\documentclass[journal]{IEEEtran}

\usepackage{amsmath, amssymb, graphicx, upgreek, cite, marginnote, color}
\usepackage{epsfig}
\usepackage[ruled, vlined,]{algorithm2e}
\usepackage{enumitem}
\newlength\myindent
\ifCLASSINFOpdf
\else
\fi
\begin{document}

\title{Optimal Experiment Design for Magnetic Resonance Fingerprinting: Cram\'er-Rao Bound Meets Spin Dynamics}
\author{Bo Zhao, \IEEEmembership{Member, IEEE},
        Justin P. Haldar, \IEEEmembership{Senior Member, IEEE},
        Congyu Liao,
        Dan Ma,
        Yun Jiang,\\
        Mark A. Griswold,
        Kawin Setsompop,
        and Lawrence L. Wald, \IEEEmembership{Member, IEEE}
\thanks{Copyright (c) 2017 IEEE. Personal use of this material is permitted. However, permission to use this material for any other purposes must be obtained from the IEEE by sending a request to pubs-permissions@ieee.org}
\thanks{This work was supported, in part, by the National Institutes of Health under Grant No. NIH-R01-EB017219, NIH-R01-EB017337, and NIH-P41-EB015896, and the National Science Foundation under Grant No. NSF-CCF-135063. B. Zhao was also supported by a Ruth L. Kirschstein National Research Service Award Postdoctoral Fellowship from the National Institutes of Health under Grant No. NIH-F32-EB024381.}
\thanks{B. Zhao is with the Athinoula A. Martinos Center for Biomedical Imaging, Massachusetts General Hospital, Charlestown, MA 02129 USA, and also with the Department of Radiology, Harvard Medical School, Boston, MA 02115 USA (email: bzhao4@mgh.harvard.edu).}
\thanks{J. P. Haldar is with the Signal and Image Processing Institute and Ming Hsieh Department of Electrical Engineering, University of Southern California, Los Angeles, CA 90089 USA (jhaldar@usc.edu).}
\thanks{C. Liao is with the Athinoula A. Martinos Center for Biomedical Imaging, Massachusetts General Hospital, Charlestown, MA 02129 USA, and also with the Department of Biomedical Engineering, Zhejiang University, Hangzhou, Zhejiang Province 310027 China (email: cliao2@mgh.harvard.edu).}
\thanks{D. Ma, Y. Jiang, and M. A. Griswold are with the Department of Radiology, Case Western Reserve University, Cleveland, OH 44106 USA (email: dan.ma@case.edu, yun.jiang@case.edu, mark.griswold@case.edu).}
\thanks{K. Setsompop and L. L. Wald are with the Athinoula A. Martinos Center for Biomedical Imaging, Massachusetts General Hospital, Charlestown, MA 02129 USA, and also with the Department of Radiology, Harvard Medical School, Boston, MA, 02115 USA, and also with the Harvard-MIT Division of Health Sciences and Technology, Massachusetts Institute of Technology, Cambridge, MA 02139 USA (email: kawin@nmr.mgh.harvard.edu, wald@nmr.mgh.harvard.edu).}
}
\markboth{IEEE Transactions on Medical Imaging}
{Shell \MakeLowercase{\textit{et al.}}: Optimal Experiment Design for Magnetic Resonance Fingerprinting}

\maketitle

\begin{abstract}
Magnetic resonance (MR) fingerprinting is a new quantitative imaging paradigm, which simultaneously acquires multiple MR tissue parameter maps in a single experiment. In this paper, we present an estimation-theoretic framework to perform experiment design for MR fingerprinting. Specifically, we describe a discrete-time dynamic system to model spin dynamics, and derive an estimation-theoretic bound, i.e., the Cram\'er-Rao bound (CRB), to characterize the signal-to-noise ratio (SNR) efficiency of an MR fingerprinting experiment. We then formulate an optimal experiment design problem, which determines a sequence of acquisition parameters to encode MR tissue parameters with the maximal SNR efficiency, while respecting the physical constraints and other constraints from the image decoding/reconstruction process. We evaluate the performance of the proposed approach with numerical simulations, phantom experiments, and in vivo experiments. We demonstrate that the optimized experiments substantially reduce data acquisition time and/or improve parameter estimation. For example, the optimized experiments achieve about a factor of two improvement in the accuracy of $T_2$ maps, while keeping similar or slightly better accuracy of $T_1$ maps. Finally, as a remarkable observation, we find that the sequence of optimized acquisition parameters appears to be highly structured rather than randomly/pseudo-randomly varying as is prescribed in the conventional MR fingerprinting experiments.
\end{abstract}


\IEEEpeerreviewmaketitle

\section{Introduction}
\IEEEPARstart{M}{agnetic} resonance (MR) fingerprinting is a novel quantitative magnetic resonance imaging (MRI) framework \cite{Ma2013}. It enables simultaneous acquisition of multiple MR tissue parameters (e.g., $T_1$, $T_2$, and spin density) in a single imaging experiment. Compared to conventional MR relaxometry techniques (e.g., \cite{Klaus2001, Deoni2003, Schmitt2004, Warntjes2007}), the original proof-of-principle MR fingerprinting implementation featured several key innovations in its encoding and decoding processes \cite{Ma2013}. Specifically, during encoding, it applies a series of time-varying, random or quasi-random data acquisition parameters (e.g., flip angles and repetition times) to probe the spin system. This generates unique transient-state signal evolutions, or fingerprints, for different MR tissue parameters. It further applies incoherent spatial encoding (e.g., variable density spiral acquisition) to collect k-space data. During decoding, it performs a simple gridding reconstruction to reconstruct contrast-weighted images, after which it applies a dictionary-based pattern matching to obtain MR tissue parameter maps of interest.

Given its many departures from conventional quantitative imaging, some fundamental questions still remain unclear about the mechanism of MR fingerprinting. For example, from a theoretical perspective, the optimality of the encoding and decoding processes for MR fingerprinting has not been examined in \cite{Ma2013}. While a recent paper has performed a theoretical analysis of the MR fingerprinting problem from a low-dimensional manifold recovery viewpoint \cite{Davies2014}, this analysis is asymptotic and probabilistic in nature, and does not provide results that could be used to ensure the quality or guide the design of finite-duration MR fingerprinting experiments.

From a practical perspective, while existing MR fingerprinting methods work well in some scenarios, there are other important cases where the performance of these methods can be much worse. For example, the accuracy of $T_2$ maps from MR fingerprinting often depends critically on the length of data acquisition, and is much worse than that of $T_1$ maps, especially with short acquisition lengths \cite{zhao2015, zhao2016, Eric2016, Cao2017, zhao2017}. In these settings, it would be highly desirable to either improve parameter mapping quality without increasing experiment duration, or reduce experiment duration without compromising parameter mapping quality.

Recently, we introduced a novel statistical imaging framework for MR fingerprinting \cite{zhao2015, zhao2016}. On the decoding side, we proposed a maximum likelihood (ML) approach that directly reconstructs MR tissue parameter maps from highly-undersampled, noisy $\mathbf{k}$-space data. Moreover, we showed analytically that the conventional reconstruction approach \cite{Ma2013} is sub-optimal from a statistical estimation perspective \cite{zhao2015, zhao2016}. The use of ML reconstruction has dramatically improved the estimation accuracy and/or reduced acquisition time. Apart from the quality of the decoding schemes, the performance of MR fingerprinting inherently depends on the quality of the available data. This motivates the optimization of MR fingerprinting on the encoding side, which naturally falls into the domain of optimal experiment design \cite{Pazman1986, Pukelsheim1993} in the statistical imaging framework.

In this paper, we address the optimal experiment design problem for MR fingerprinting. Our goal is to encode MR tissue parameters into the most informative measurements in the presence of noise. Specifically, we first model spin dynamics with a discrete-time dynamic system. We then calculate an estimation-theoretic bound, i.e., the Cram\'er-Rao bound (CRB) \cite{kay1993}, as a measure of the signal-to-noise (SNR) efficiency for MR fingerprinting experiments. We further utilize this bound to formulate an optimal experiment design problem to choose MR fingerprinting acquisition parameters for the maximal SNR efficiency, while respecting both physical constraints and other constraints from image decoding. We show representative results from numerical simulations, phantom experiments, and in vivo experiments to illustrate the performance of the proposed framework. As a remarkable observation, we find that the sequence of optimized acquisition parameters appears to be highly structured rather than randomly/pseudo-randomly varying as used in the existing MR fingerprinting experiments (e.g., \cite{Ma2013, Jiang2015, Jakob2017, Jiang2017, Benedikt2016}).

A preliminary account of this work was presented in our early conference papers \cite{zhao2016c, zhao2016b, zhao2018}, which, to the best of our knowledge, first introduced the use of estimation-theoretic bounds for MR fingerprinting experiment design. Recently, similar problems have also been investigated by other researchers. For example, Assl\"{a}nder et al. applied our CRB-based experiment design framework to a new MRF imaging sequence \cite{Jakob2017} under the polar coordinates of the Bloch equation \cite{Asslander2017}. Maidens et al. provided an optimal control interpretation of the experiment design problem, for which they developed a dynamic programming based algorithm \cite{Maidens2016}. However, due to the curse of dimensionality for dynamic programming \cite{Bertsekas2012}, the feasibility of their approach was only demonstrated in a highly simplified scenario that does not reflect the full complexities of real MRF applications. Besides CRB-based approaches, the experiment design problem was also addressed from other perspectives. For example, Cohen et al. optimized MR fingerprinting acquisition parameters by maximizing the discrimination power between different tissue types \cite{cohen2017}.

Note that the CRB has been previously used in analyzing and designing experiments for conventional MR relaxometry (e.g., \cite{jone1996, Zhang1998, Haldar2009, Funai2010, zhao2014, Akaakaya2015, Christina2016, Nataraj2017, Teixeira2017}). Here we extend these approaches to MR fingerprinting, which is a unique transient-state quantitative imaging technique that encodes MR tissue parameters into spin dynamics. In this context, the CRB calculation can be more complex, which requires handling a dynamic system, rather than an analytical signal model as in conventional MR relaxometry \cite{jone1996, Zhang1998, Haldar2009, Funai2010, zhao2014, Akaakaya2015, Christina2016, Nataraj2017, Teixeira2017}. In this paper, we introduce an efficient way of calculating the CRB for dynamic systems described by the Bloch equation. In particular, we show that the CRB calculation can be done by iterating a set of difference equations.

For easy reference, we summarize here the key notations and symbols used in the paper. We use $\mathbb{R}$ to denote the field of real numbers, and use $\mathbb{R}^{n}$ and $\mathbb{R}^{m\times n}$ to respectively denote the space of real vectors of length $n$ and the space of real $m\times n$ matrices. We use bold letters (e.g., $\mathbf{x}$ or $\mathbf{X}$) to denote vectors or matrices. We respectively use $\mathbf{X}^T$, $\mathbf{X}^{-1}$, and $\mathrm{tr}\left(\mathbf{X}\right)$ to denote the transpose, inverse, and trace of $\mathbf{X}$. For a scalar-valued function $f: \mathbb{R}^{N} \rightarrow \mathbb{R}$ with the argument $\mathbf{x} \in \mathbb{R}^N$, we define its gradient, i.e., $\partial{f}/\partial{\mathbf{x}}$, as an $N \times 1$ vector with $\left[\partial{f}/\partial{\mathbf{x}}\right]_n = \partial{f}/\partial{\mathbf{x}_n}$; for a vector-valued function $F: \mathbb{R}^{N} \rightarrow \mathbb{R}^{M}$ with the argument $\mathbf{x} \in \mathbb{R}^N$, we define its Jacobian matrix, i.e., $\partial{F}/\partial{\mathbf{x}}$, as an $M \times N$ matrix with $\left[\partial{F}/\partial{\mathbf{x}} \right]_{m, n}= \partial{F_m}/\partial{\mathbf{x}_n}$; and for a matrix-valued function $\boldsymbol{\mathit{F}}: \mathbb{R} \rightarrow \mathbb{R}^{M\times N}$ with the argument $x \in \mathbb{R}$, we define its derivative as an $M\times N$ matrix with $\left[\partial\boldsymbol{\mathit{F}}/\partial{x} \right]_{m, n}= \partial{\boldsymbol{\mathit{F}}_{m, n}}/\partial{x}$. For a random vector $\mathbf{x} \in \mathbb{R}^{N}$, we denote its expectation as $\mathbb{E}\left[\mathbf{x}\right] \in \mathbb{R}^{N}$, and its covariance matrix as $\mathrm{Cov}\left(\mathbf{x}\right) = \mathbb{E}\left[\left(\mathbf{x} - \mathbb{E}\left(\mathbf{x}\right)\right)\left(\mathbf{x} - \mathbb{E}\left(\mathbf{x}\right)\right)^T\right] \in \mathbb{R}^{N \times N}$.

The rest of the paper is organized as follows. Section \ref{sec:method} presents the proposed framework in detail, which starts with a state-space model for spin dynamics, followed by the calculation of the Cram\'er-Rao bound and the optimal design of MR fingerprinting experiments. Section \ref{sec:results} demonstrates the performance of the proposed approach with numerical simulations, phantom experiments, and in vivo experiments. Section \ref{sec:discussion} discusses the related issues and future work, followed by the concluding remarks in Section \ref{sec:conclusion}.

\section{Proposed Framework}
\label{sec:method}

\subsection{Signal Model}
A number of MR fingerprinting imaging sequences have recently been developed (e.g., \cite{Ma2013, Jiang2015, Jakob2017, Benedikt2016, Jiang2017}). In this subsection, we start by formalizing a generic state-space model \cite{Luenberger1979} for spin dynamics that underlie all MR fingerprinting sequences, and then give an example by specializing this model to a representative MR fingerprinting sequence.
\subsubsection{State-Space Model}
Spin dynamics are governed by the Bloch equation \cite{Bloch1946}, which is a system of first-order ordinary differential equations. While the magnetization evolves in continuous time, we are mainly interested in its values at a finite set of time points in an MR fingerprinting experiment. Here we consider a discrete-time state-space model for simplicity, which completely captures the features of interest for spin dynamics in continuous time. Moreover, we focus on the magnetization evolution with respect to a small sample of tissue (i.e., a voxel); other issues related to spatial encoding will be discussed later.

In the presence of magnetic field inhomogeneity (e.g., with the use of gradients), intravoxel spin dephasing occurs. To account for this effect, we model multiple isochromats in a voxel, each of which  represents an ensemble of spins that have the same resonance frequency \cite{Liang1999}.  Here we use the isochromat-summation approach \cite{Pavel1997, Malik2016} to approximate the magnetization evolution for the voxel of interest. This approach performs Bloch simulations for each individual isochromat, and then sums up their magnetization evolutions to obtain the ensemble magnetization evolution for a voxel. From a dynamic system viewpoint, this approach essentially employs an independent state equation to describe the magnetization evolution of each isochromat, and then obtains the ensemble magnetization through an observation equation.

In the following, we start by describing the state equation for a single isochromat. Here we divide the voxel of interest, i.e., $\bigtriangleup$, into a set of sufficiently small, equal-sized subvoxels $\left\{\bigtriangleup_{\mathbf{r}}\right\}$, where $\bigtriangleup_{\mathbf{r}}$ is centered at $\mathbf{r}$ and $\bigtriangleup_{\mathbf{r}} \subset \bigtriangleup$. We assume that there exists a single isochromat for each $\bigtriangleup_{\mathbf{r}}$.  Let $\mathbf{M}_{\mathbf{r}}[n] \in \mathbb{R}^3$ denote the magnetization associated with the isochromat at $\bigtriangleup_{\mathbf{r}}$ at the end of the $n$th repetition time (immediately before the next signal excitation). The magnetization evolution can be described by the following state equation:
\begin{equation}
\label{eq:signal_evolution}
\mathbf{M}_{\mathbf{r}}[n] = \mathbf{A}_{\mathbf{r}}(\mathbf{u}[n], \boldsymbol{\uptheta})\mathbf{M}_{\mathbf{r}}[n-1] + \mathbf{B}_{\mathbf{r}}(\mathbf{u}[n], \boldsymbol{\uptheta}),
\end{equation}
for $n = 1, \cdots, N$, where $\boldsymbol{\uptheta} \in \mathbb{R}^p$ contains the unknown parameters in an MR fingerprinting experiment, including the tissue-specific parameters (e.g., $T_1$, $T_2$, and spin density) and experiment-specific parameters (e.g., off-resonance frequency); $\mathbf{u}[n] \in \mathbb{R}^{q}$ contains the data acquisition parameters applied during the $n$th repetition time, including the flip angle $\alpha_n$, the phase of the radio frequency (RF) pulse $\phi_n$, the echo time $TE_n$, and the repetition time $TR_n$; $\mathbf{A}_{\mathbf{r}}(\mathbf{u}[n], \boldsymbol{\uptheta}) \in \mathbb{R}^{3\times 3}$ and $\mathbf{B}_{\mathbf{r}}(\mathbf{u}[n], \boldsymbol{\uptheta}) \in \mathbb{R}^{3\times 1}$ respectively denote the system matrix and input matrix for the $n$th repetition time. Note that in \eqref{eq:signal_evolution}, we implicitly assume that all the subvoxels share the same set of parameters $\boldsymbol{\uptheta}$. Moreover, assuming that the imaging experiment starts from thermal equilibirium, the initial condition for \eqref{eq:signal_evolution} is given by $\mathbf{M}_{\mathbf{r}}[0] = \left[0, 0, M_0(\mathbf{r})\right]^T$, where $M_0(\mathbf{r}) = M_0/N_v$, $M_0$ denotes the magnitude of the magnetization for the voxel $\bigtriangleup$ at thermal equilibrium, and $N_v$ denotes the number of sub-voxels in $\bigtriangleup$.

Next, we describe the observation equation. Note that the magnetization is measured at the corresponding echo time (i.e., $TE_n$) after each RF excitation, and the signal detected by the receiver coil is proportional to the transverse component of the magnetization. Denoting $\mathbf{m}_\mathbf{r}[n] \in \mathbb{R}^2$ as the transverse magnetization associated with the isochromat at $\bigtriangleup_{\mathbf{r}}$ at the $n$th echo time, we have
\begin{equation}
\label{eq:obs_model}
\mathbf{m}_\mathbf{r}[n] = \mathbf{C}_{\mathbf{r}}(\mathbf{u}[n], \boldsymbol{\uptheta})\mathbf{M}_{\mathbf{r}}[n-1],
\end{equation}
for $n = 1, \cdots, N$, where $\mathbf{C}_{\mathbf{r}}(\mathbf{u}[n], \boldsymbol{\uptheta}) \in \mathbb{R}^{2 \times 3}$ denotes the output matrix. Summing up the magnetizations of all the isochromats, we can obtain the transverse magnetization for $\bigtriangleup$ as
\begin{equation}
\label{eq:obs_model1}
\mathbf{m}[n] = \sum_{\mathbf{r}: \bigtriangleup_{\mathbf{r}} \subset \bigtriangleup} \mathbf{m}_\mathbf{r}[n].
\end{equation}
Putting together \eqref{eq:obs_model} and \eqref{eq:obs_model1}, we have
\begin{equation}
\label{eq:obs_model2}
\mathbf{m}[n] =  \sum_{\mathbf{r}: \bigtriangleup_{\mathbf{r}} \subset \bigtriangleup}\mathbf{C}_{\mathbf{r}}(\mathbf{u}[n], \boldsymbol{\uptheta})\mathbf{M}_{\mathbf{r}}[n-1].
\end{equation}

The equations \eqref{eq:signal_evolution} and \eqref{eq:obs_model2} together form a state-space model, which can describe spin dynamics for various MR fingerprinting imaging sequences (e.g., \cite{Ma2013, Jiang2015, Jakob2017, Benedikt2016}). Note that this model is nonlinear and time-varying in nature.

\subsubsection{Example Sequence}
As an example, we illustrate the above state-space model using a widely-used MR fingerprinting sequence, i.e., inversion recovery fast imaging with steady-state precession (IR-FISP) sequence \cite{Jiang2015}. This sequence is robust to off-resonance effects with the use of spoiler gradients \cite{Jiang2015}. Here the unknown parameters are $\boldsymbol{\uptheta} = [T_1, T_2, M_0]^T$, and the acquisition parameters are $\mathbf{u}[n] = [\alpha_n, \phi_n, TE_n, TR_n]^T$.

In the IR-FISP sequence, three physical processes drive the magnetization evolutions: (1) RF excitation; (2) spin relaxation; and (3) spin dephasing. In the following, we use the above isochromat-summation approach to model spin dynamics. Specifically, under Cartesian coordinates in the rotating frame \cite{Liang1999},\footnote{Alternatively, the model can be described in other coordinate systems (e.g., the polar coordinates \cite{Asslander2017}); however, note that the Cram\'er-Rao bound does not change under any transform of the coordinate system for the Bloch equation.} we can form the system matrix $\mathbf{A}_{\mathbf{r}}$ as
\begin{equation}
\label{eq:system_matrix}
\mathbf{A}_{\mathbf{r}}(\mathbf{u}[n], \boldsymbol{\uptheta}) = \mathbf{G}(\beta_{\mathbf{r}})\mathbf{R}(T_1, T_2, TR_n)\mathbf{Q}(\alpha_n, \phi_n),
\end{equation}
where $\mathbf{Q}(\alpha_n, \phi_n) \in \mathbb{R}^{3 \times 3}$ models the RF excitation, i.e.,
\begin{align}
&\mathbf{Q}(\alpha_n, \phi_n) =
\begin{bmatrix}
\cos(\phi_n)& \sin(\phi_n) &0\\
-\sin(\phi_n)& \cos(\phi_n)& 0\\
0& 0& 1
\end{bmatrix} \nonumber\\
&\begin{bmatrix}
1& 0 &0\\
0 & \cos(\alpha_n)& \sin(\alpha_n)\\
0& -\sin(\alpha_n) & \cos(\alpha_n)
\end{bmatrix}
\begin{bmatrix}
\cos(\phi_n)& -\sin(\phi_n) &0\\
\sin(\phi_n)& \cos(\phi_n)& 0\\
0& 0& 1
\end{bmatrix};\nonumber
\end{align}
$\mathbf{R}(T_1, T_2, t) \in \mathbb{R}^{3\times 3}$ models the spin relaxation, i.e.,
\begin{equation}
\mathbf{R}(T_1, T_2, t) =
\begin{bmatrix}
e^{-t/T_2} & 0 & 0\\
0 & e^{-t/T_2} & 0 \\
0 & 0 & e^{-t/T_1}
\end{bmatrix}; \nonumber
\end{equation}
and $\mathbf{G}(\beta_\mathbf{r}) \in \mathbb{R}^{3\times 3}$ models the spin dephasing, i.e.,
\begin{equation}
\mathbf{G}(\beta_\mathbf{r}) =
\begin{bmatrix}
\cos(\beta_\mathbf{r})& \sin(\beta_\mathbf{r}) & 0\\
-\sin(\beta_\mathbf{r}) & \cos(\beta_\mathbf{r}) &0\\
0 & 0 & 1
\end{bmatrix},
\nonumber
\end{equation}
where $\beta_{\mathbf{r}}$ denotes the phase dispersion associated with the isochromat at $\bigtriangleup_{\mathbf{r}}$. Note that in the IR-FISP sequence, $\beta_{\mathbf{r}}$ is dominated by the impact of spoiler gradients, although there are various other factors (e.g., diffusion effects) that can also contribute to spin dephasing.

Moreover, we can form the input matrix $\mathbf{B}_{\mathbf{r}}$ as follows:
\begin{equation}
\label{eq:input_matrix}
\mathbf{B}_{\mathbf{r}}(\mathbf{u}[n], \boldsymbol{\uptheta}) =  M_0(\mathbf{r})\mathbf{b}(T_1,  TR_n),
\end{equation}
where $\mathbf{b}(T_1, t) = \left[0, 0, 1 - e^{-t/T_1}\right]^T$. Note that $\mathbf{B}_{\mathbf{r}}(\mathbf{u}[n], \boldsymbol{\uptheta})$ models the recovery of the longitudinal magnetization.

Finally, we can form the output matrix $\mathbf{C}_{\mathbf{r}}$ as
\begin{equation}
\label{eq:output_matrix}
\mathbf{C}_{\mathbf{r}}(\mathbf{u}[n], \boldsymbol{\uptheta}) = \mathbf{P}\mathbf{R}(T_1, T_2, TE_n)\mathbf{Q}(\alpha_n, \phi_n),
\end{equation}
where $\mathbf{P} \in \mathbb{R}^{2 \times 3}$ is the projection matrix that extracts the transverse magnetization, i.e.,
\begin{equation}
\mathbf{P} =
\begin{bmatrix}
1& 0& 0\\
0& 1& 0
\end{bmatrix}, \nonumber
\end{equation}
and $\mathbf{R}(T_1, T_2, TE_n)$ and $\mathbf{Q}(\alpha_n, \phi_n)$ are defined in the same way as before. Here $\mathbf{C}_{\mathbf{r}}(\mathbf{u}[n], \boldsymbol{\uptheta})$ models the RF excitation as well as the spin relaxation. Note that this matrix does not account for spin dephasing, since a spoiler gradient is placed after the echo time within each repetition time.

\subsection{Cram\'er-Rao Bound}
We proceed to describe the data model, with which we calculate the CRB and perform experiment design. Note that the spin dynamics described above are utilized to perform contrast encoding in MR fingerprinting experiments. Besides contrast encoding, the data generating process also encompasses spatial encoding and noise contamination. In \cite{zhao2015, zhao2016}, we described a data model for this data generating process. In principle, we can use it to calculate the CRB and perform experiment design. In practice, however, this is often computationally very expensive with the use of the non-Cartesian Fourier transform in spatial encoding. Here we describe a practical approach, in which we ignore the spatial encoding, and use the following simplified data model:
\begin{equation}
\label{eq:data_model}
\mathbf{s}[n] = \mathbf{m}[n] + \mathbf{z}[n],
\end{equation}
for $n = 1, \cdots, N$. Here $\mathbf{s}[n] \in \mathbb{R}^2$ is a vector that contains the magnetizations collected at the $n$th echo time, and $\left\{\mathbf{z}[n]\right\}_{n=1}^{N}$ denotes independent, identically distributed Gaussian noise with $\mathbf{z}[n] \sim N(\mathbf{0}, \sigma^2\mathbf{I})$. Note that \eqref{eq:data_model} corresponds to the data model for a single voxel nuclear magnetic resonance (NMR) experiment. Alternatively, it can also be viewed as the data model for a fully-sampled imaging experiment,\footnote{More precisely, this is equivalent to a single-channel Nyquist-sampled Cartesian Fourier acquisition with a discrete Fourier transform based image reconstruction.} in which there is no ``crosstalk'' between magnetization evolutions at different voxels. Despite such simplification, we will demonstrate later that the experiment design with \eqref{eq:data_model} can be very effective for highly-undersampled MR fingerprinting experiments.

Next, we derive the CRB for the data model \eqref{eq:data_model}. From estimation theory, the CRB provides a lower bound on the covariance of any unbiased estimator under mild regularity conditions, and this bound can be asymptotically achieved by the ML estimator \cite{kay1993}. Mathematically, the CRB can be expressed as the following information inequality \cite{kay1993}:
\begin{equation}
\label{eq:CRB}
\mathbb{E}\left\{\left(\boldsymbol{\uptheta} - \hat{\boldsymbol{\uptheta}}\right)\left(\boldsymbol{\uptheta} - \hat{\boldsymbol{\uptheta}}\right)^T\right\} \geq \mathbf{V}(\boldsymbol{\uptheta}) = \mathbf{I}^{-1}\left(\boldsymbol{\uptheta}\right),
\end{equation}
for any unbiased estimator $\hat{\boldsymbol{\uptheta}}$, where $\mathbf{I}(\boldsymbol{\uptheta}) \in \mathbb{R}^{p \times p}$ denotes the Fisher information matrix (FIM) defined as
\begin{eqnarray}
\label{eq:FIM}
\mathbf{I}(\boldsymbol{\uptheta}) &=& \mathbb{E}\left[\left(\frac{\partial \ln{p(\{\mathbf{s}[n]\}; \boldsymbol{\uptheta}})}{\partial \boldsymbol{\uptheta}}\right)\left(\frac{\partial \ln{p(\{\mathbf{s}[n]\}; \boldsymbol{\uptheta}})}{\partial \boldsymbol{\uptheta}}\right)^T\right], \nonumber
\end{eqnarray}
$\mathbf{V}(\boldsymbol{\uptheta}) \in \mathbb{R}^{p \times p}$ denotes the CRB matrix, and $\ln{p(\mathbf{x}; \boldsymbol{\uptheta})}$ denotes the log-likelihood function of the observation $\mathbf{x}$ parameterized by $\boldsymbol{\uptheta}$. In \eqref{eq:CRB}, the matrix inequality $\mathbf{A}\geq\mathbf{B}$ means that $\mathbf{A} - \mathbf{B}$ is a positive semidefinite matrix. Note that both the CRB and FIM depend on the underlying tissue parameter $\boldsymbol{\uptheta}$, given that the data model \eqref{eq:data_model} is nonlinear with respect to $\boldsymbol{\uptheta}$. Moreover, we can obtain the bound on the variance of individual tissue parameter estimate by extracting the corresponding diagonal entry of the CRB matrix, i.e.,
\begin{equation}
\label{eq:CRB_voxel}
\mathrm{Var}\left(\hat{\boldsymbol{\uptheta}}_i\right) \geq \left[\mathbf{V}(\boldsymbol{\uptheta})\right]_{i, i}.
\end{equation}

To calculate the CRB in \eqref{eq:CRB}, we need to compute the FIM $\mathbf{I}\left(\boldsymbol{\uptheta}\right)$. For the additive Gaussian data model in \eqref{eq:data_model}, the FIM has the particularly simple form, which can be written as follows \cite{kay1993}:
\begin{equation}
\label{eq:FIM_final}
\mathbf{I}\left(\boldsymbol{\uptheta}\right) = \frac{1}{\sigma^2} \sum_{n=1}^N\mathbf{J}^{T}_n\left(\boldsymbol{\uptheta}\right)\mathbf{J}_n\left(\boldsymbol{\uptheta}\right),
\end{equation}
where $\mathbf{J}_n\left(\boldsymbol{\uptheta}\right) = \partial{\mathbf{m}}[n]/\partial{\boldsymbol{\uptheta}} \in \mathbb{R}^{2 \times p}$ is the Jacobian matrix.

Finally, we describe the calculation of $\mathbf{J}_n\left(\boldsymbol{\uptheta}\right)$ for \eqref{eq:FIM_final}. Given the state-space model in \eqref{eq:signal_evolution} and \eqref{eq:obs_model2}, such calculation is equivalent to iterating a set of difference equations. More specifically, noting that
\begin{align}
\label{eq:Jacobian}
\mathbf{J}_n\left(\boldsymbol{\uptheta}\right) = \frac{\partial{\mathbf{m}[n]}}{\partial{\boldsymbol{\uptheta}}} =
\begin{bmatrix}
\frac{\partial{}}{\partial{\boldsymbol{\uptheta}_1}}\mathbf{m}[n] & \cdots & \frac{\partial{}}{\partial{\boldsymbol{\uptheta}_p}}\mathbf{m}[n]
\end{bmatrix},
\end{align}
we can compute $\partial\mathbf{m}[n]/\partial\boldsymbol{\uptheta}_i$ for each entry of $\boldsymbol{\uptheta}$. This can be done as follows. First, we take the derivative with respect to $\boldsymbol{\uptheta}_i$ on both sides of \eqref{eq:obs_model2}, which yields
\begin{multline}
\label{eq:derivative1}
\frac{\partial \mathbf{m}[n]}{\partial \boldsymbol{\uptheta}_i} =  \sum_{\mathbf{r}: \bigtriangleup_{\mathbf{r}} \subset \bigtriangleup}\frac{\partial \mathbf{C}_{{\mathbf{r}}}(\mathbf{u}[n], \boldsymbol{\uptheta})}{\partial \boldsymbol{\uptheta}_i}\mathbf{M}_{\mathbf{r}}[n-1] ~~ +  \\
  \sum_{\mathbf{r}: \bigtriangleup_{\mathbf{r}} \subset \bigtriangleup} \mathbf{C}_{\mathbf{r}}(\mathbf{u}[n], \boldsymbol{\uptheta})\frac{\partial\mathbf{M}_{\mathbf{r}}[n-1]}{\partial \boldsymbol{\uptheta}_i},
\end{multline}
for $n = 1, \cdots, N$. Then we invoke the derivative with respect to $\boldsymbol{\uptheta}_i$ on both sides of \eqref{eq:signal_evolution}, which yields
\begin{multline}
\label{eq:derivative2}
\frac{\partial \mathbf{M}_{\mathbf{r}}[n]}{\partial \boldsymbol{\uptheta}_i} = \frac{\partial \mathbf{A}_{\mathbf{r}}(\mathbf{u}[n], \boldsymbol{\uptheta})}{\partial \boldsymbol{\uptheta}_i}\mathbf{M}_{\mathbf{r}}[n-1]  \\
+ \mathbf{A}_{\mathbf{r}}(\mathbf{u}[n], \boldsymbol{\uptheta})\frac{\partial \mathbf{M}_{\mathbf{r}}[n-1]}{\partial \boldsymbol{\uptheta}_i}
+ \frac{\partial \mathbf{B}_{\mathbf{r}}(\mathbf{u}[n], \boldsymbol{\uptheta})}{\partial \boldsymbol{\uptheta}_i}.
\end{multline}
Now we can iterate the two difference equations \eqref{eq:derivative1} and \eqref{eq:derivative2} to calculate $\frac{\partial{}}{\partial{\boldsymbol{\uptheta}_i}}\mathbf{m}[n]$. Note that the initial conditions are given by $\mathbf{M}_{\mathbf{r}}[0] = \left[0, 0, M_0(\mathbf{r})\right]^T$ and $\partial\mathbf{M}_{\mathbf{r}}[0]/\partial {\boldsymbol{\uptheta}_i}= \left[0, 0, \partial M_0(\mathbf{r})/\partial{\boldsymbol{\uptheta}_i}\right]^T$. For the sake of concreteness, we illustrate the above procedure with the example IR-FISP sequence in the Appendix.

\subsection{Optimal Experiment Design}
Given that the CRB provides a lower bound on the smallest possible variance for any unbiased estimator, it can be used to characterize the SNR efficiency of an imaging experiment. This helps understand the potential reliability of an MR fingerprinting experiment, and figure out how much acquisition time is necessary to achieve a certain level of quantitative accuracy. More importantly, we can use the CRB as a principled tool to optimize the encoding process of an MR fingerprinting experiment. For example, given a set of representative MR tissue parameters $\{\boldsymbol{\uptheta}^{(l)}\}_{l=1}^L$, we could optimize the data acquisition parameters of an MR fingerprinting experiment to maximize its SNR efficiency. Mathematically, such an optimal experiment design problem can be formulated as follows\footnote{The term ``optimal experiment design'' refers to the goal of the problem formulation, i.e., maximizing the SNR efficiency of an experiment. The use of this term follows the convention in statistics \cite{Fedorov1972, Pazman1986, Pukelsheim1993} and in MR imaging \cite{Xie2008, Poot2010, Lampinen2017}, and is unrelated to whether a specific solution algorithm produces a globally optimal design or not.}:
\begin{equation}
\label{eq:general_formulation}
\begin{split}
\underset{\mathbf{u}}{\text{min}}&\sum_{l=1}^L\Psi\left(\mathbf{V}(\boldsymbol{\uptheta}^{(l)})\right)\\
&\text{s.t.}~~\mathbf{u} \subset \mathcal{U},
\end{split}
\end{equation}
where $\Psi(\cdot)$ denotes the design criterion, which is a scalar function of the CRB matrix; $\mathbf{u} = \left[\mathbf{u}[1] , \cdots, \mathbf{u}[N] \right] \in \mathbb{R}^{q\times N}$ denotes the acquisition parameters for an MR fingerprinting experiment with $N$ time points; and $\mathcal{U} \subset \mathbb{R}^{q\times N}$ denotes the constraint set for feasible data acquisition parameters.

Note that, for the sake of concrete illustration, we assume that the total number of time points, i.e., $N$, is given in \eqref{eq:general_formulation}. In practice, $N$ can be specified according to the desired experiment duration and the constraints on TRs. Moreover, given that the CRB matrix depends on the underlying tissue parameter, we assume that, for a specific application of interest (e.g., neuroimaging), we have the knowledge of the range of MR tissue parameter values prior to our experiment design. While it is desirable to design experiments that are universally optimal for all possible parameters within the range, this is often not feasible. As such, we select a few representative tissues as a practical compromise.

As an example, we specialize \eqref{eq:general_formulation} to optimize MR fingerprinting experiments with the IR-FISP sequence. First, we specify the design criterion for $\Psi(\cdot)$. Note that there are various information criteria that can be used, including the A-optimality, D-optimality, and E-optimality criteria (see \cite{Pukelsheim1993} for a comprehensive survey). Here we choose the A-optimality criterion \cite{Pukelsheim1993}, which minimizes the trace of the CRB matrix (i.e., the total variance of tissue parameter estimates). Further, we incorporate weightings into the design criterion, which is motivated by: (1) the CRBs for different tissue parameters are often at very different scales; and (2) we may want to tailor a design to the parameters that are most relevant to specific applications of interest. Accordingly, we have $\Psi(\cdot) = \mathrm{tr}\left(\mathbf{W}\mathbf{V}(\boldsymbol{\uptheta})\right)$, where $\mathbf{W}$ is a diagonal matrix whose entries contain weightings for different tissue parameters.

Second, we specify the data acquisition parameters $\mathbf{u} = \left[\mathbf{u}[1] , \cdots, \mathbf{u}[N] \right]$. Note that the acquisition parameters for the IR-FISP sequence include the flip angles, RF pulse phases, echo times, and repetition times. Following the early work \cite{Jiang2015}, we assume the flip angles and repetition times to be the design parameters, while fixing the RF pulse phases and echo times. Accordingly, we have $\mathbf{u}[n] = [\alpha_n, TR_n]^T$, for $n = 1, \cdots, N$.

Lastly, we specify the constraint set $\mathcal{U}$ for the acquisition parameters. Taking into account various physical considerations (e.g., specific absorption rate (SAR), and/or total acquisition time), we impose upper bounds and lower bounds for the acquisition parameters, i.e., $TR_n \in \left[TR_n^{\text{min}}, TR_n^{\text{max}}\right]$,\footnote{Alternatively, we could replace this constraint with a constraint on the total acquisition time as in our previous work \cite{zhao2016b, zhao2016c}. In principle, such a formulation could allow larger variations of TRs, which could potentially improve the CRB of the optimized experiments. However, note that this formulation has to simultaneously determine both $N$ and $\{\alpha_n, TR_n\}_{n = 1}^N$, which often leads to a highly non-convex (or mixed integer) optimization problem. Without loss of generality, we have decided to focus this paper on a simpler optimization formulation that avoids this issue.} and $\alpha_n \in \left[\alpha_n^{\text{min}}, \alpha_n^{\text{max}}\right]$, for $n = 1, \cdots, N$. Accordingly, we can formulate the optimal experiment design problem as follows:
\begin{eqnarray}
\label{eq:exp_design}
& \underset{\{\alpha_n, TR_n\}}{\text{min}} \sum_{l = 1}^L \mathrm{tr}  \left(\mathbf{W}\mathbf{V}\left(\boldsymbol{\uptheta}^{(l)}\right)\right) \nonumber\\
\text{s.t.}&~~ TR_n^{\text{min}} \leq TR_n \leq TR_n^{\text{max}}, ~~ 1\leq n \leq N, \nonumber \\
&~~~~~ \alpha_n^{\text{min}} \leq \alpha_n \leq \alpha_n^{\text{max}}, ~~~~~ 1\leq n \leq N.
\end{eqnarray}

Besides the physical constraints, it is often useful to take into account other constraints from image decoding/reconstruction, especially when dealing with highly-undersampled MR fingerprinting experiments. While different reconstruction methods may use different strategies for image decoding (from highly-undersampled data), they often benefit from magnetization evolutions being smoothly varying.\footnote{For example, for the conventional reconstruction method \cite{Ma2013} that utilizes direct pattern matching, smoother magnetization evolutions often have less correlation with noise-like aliasing artifacts. This causes the underlying magnetization evolution to be better differentiated from aliasing artifacts, leading to improved accuracy in dictionary matching. As another example, for the low-rank/subspace reconstruction method \cite{zhao2015c, zhao2015b, zhao2017},  smooth magnetization evolutions often result in a smaller low-rank approximation error given the same rank value. This in turn enables better reconstruction accuracy. Finally, for statistical reconstruction methods \cite{zhao2016, zhao2015b} that involve solving nonconvex optimization problems, an initialization from the improved pattern-matching reconstruction or low-rank reconstruction method often leads to better performance.} There are a number of ways to enforce this property. Here we incorporate an additional set of constraints into \eqref{eq:exp_design}, which restricts the maximum flip angle variations between consecutive time points. As will be demonstrated later, such constraints are effective in promoting smooth magnetization evolutions, which can yield better performance for image reconstruction with highly-undersampled data. Accordingly, we reformulate the optimal experiment design problem as follows:
\begin{eqnarray}
\label{eq:exp_design2}
& \underset{\left\{\alpha_n, TR_n\right\}_{n = 1}^N}{\text{min}} \sum_{l = 1}^L \mathrm{tr}  \left(\mathbf{W}\mathbf{V}(\boldsymbol{\uptheta}^{(l)})\right) \nonumber\\
~~\text{s.t.}& TR_n^{\text{min}} \leq TR_n \leq TR_n^{\text{max}},~ 1\leq n \leq N, \nonumber \\
& ~~~ \alpha_n^{\text{min}} \leq \alpha_n \leq \alpha_n^{\text{max}}, ~~~~ 1\leq n \leq N, \nonumber \\
&~~~~|\alpha_{n+1} - \alpha_{n}| \leq \Delta\alpha_n^{\text{max}}, ~~~ 1\leq n \leq N - 1.
\end{eqnarray}
where $\Delta\alpha_n^{\text{max}}$ specifies the maximum flip angle variations between $\alpha_n$ and $\alpha_{n+1}$.

The proposed formulations in \eqref{eq:exp_design} and \eqref{eq:exp_design2} result in nonlinear and nonconvex optimization problems. A number of numerical algorithms can be employed to solve the optimization problems, including standard nonlinear optimization methods \cite{nocedal2006} and stochastic optimization methods \cite{Spall2003}. As an example, we use a state-of-the-art nonlinear optimization method, i.e., sequential quadratic programming (SQP) \cite{nocedal2006}, to seek local minima for \eqref{eq:exp_design} and \eqref{eq:exp_design2}. The SQP algorithm is an iterative algorithm, and at each iteration, it performs a quadratic approximation of the cost function at the current solution, and also linearizes the constraints. Then it solves a constrained quadratic optimization problem to update the solution. As with other nonlinear optimization methods, the performance of the SQP algorithm is generally dependent on initialization. Here we initialize the algorithm with the acquisition parameters from the conventional MR fingerprinting experiment \cite{Jiang2015}, with which the algorithm consistently yields good performance, although other initialization schemes (e.g., a multi-start strategy) may lead to better performance.
\begin{figure*}[!htb]
\centering
{\includegraphics{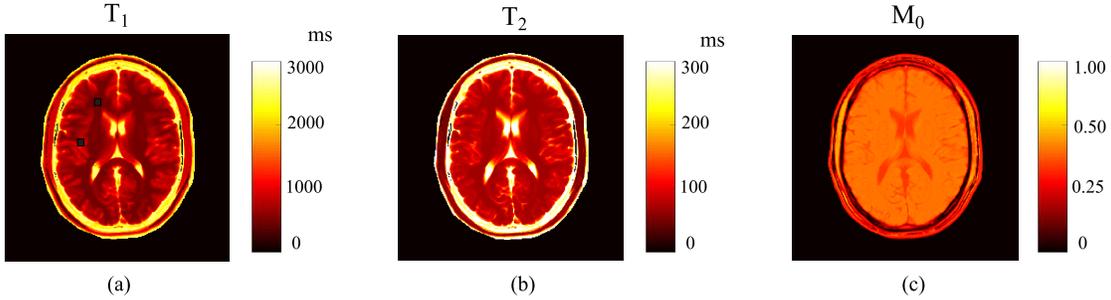}}
\caption{Ground truth parameter maps for the brain phantom: (a) $T_1$ map, (b) $T_2$ map, and (c) $M_0$ map. Note that the two ROIs (respectively in the white matter and gray matter) are marked in (a).}
\label{fig:ground_truth}
\end{figure*}
\section{Results}
\label{sec:results}
In this section, we show representative results from numerical simulations, phantom experiments, and in vivo experiments to illustrate the performance of the proposed approach.
\subsection{Simulations}

\subsubsection{General Setup}
We created a numerical brain phantom to simulate single-channel MR fingerprinting experiments. We took the $T_1$, $T_2$, and $M_0$ maps from the Brainweb database \cite{Collins1998} as the ground truth, as shown in Fig.~\ref{fig:ground_truth}. We set the experimental field-of-view (FOV) as $300 \times 300~\mathrm{mm}^2$, and the matrix size as $256 \times 256$. We simulated MR fingerprinting experiments with the IR-FISP sequence \cite{Jiang2015}, which is robust to main magnetic field inhomogeneity. Moreover, for simplicity, we assumed that the transmit RF field was homogeneous in the simulations.
\begin{figure}[!htb]
\centering
{\includegraphics{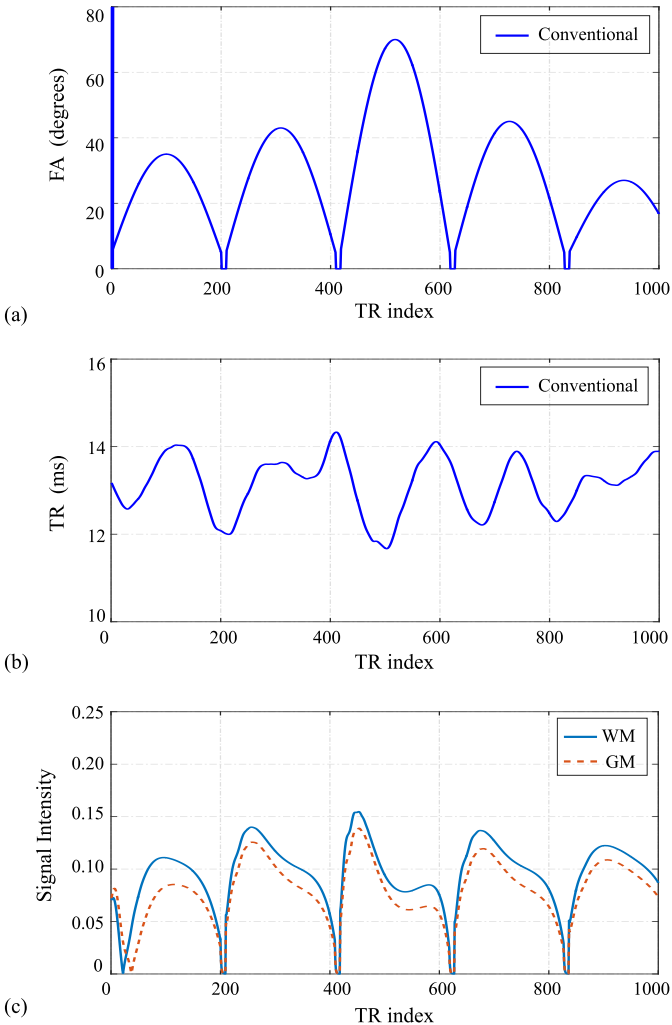}}
\caption{Acquisition parameters of the conventional scheme with $N = 1000$ as well as the resulting magnetization evolutions. (a) Flip angle train. (b) Repetition time train. (c) Magnetization evolutions for the white matter and gray matter ROIs marked in Fig.~\ref{fig:ground_truth} (a). Note that the first RF pulse is $180^{\circ}$, which exceeds the scale of the vertical axes in (a).}
\label{fig:conv_acq}
\end{figure}

We performed Bloch simulations to generate contrast-weighted images. Here we used the isochromat-summation approach \cite{Pavel1997, Malik2016}, in which we simulated magnetization evolutions with 400 isochromats for each voxel. In Bloch simulations, we considered three different sets of acquisition parameters: (1) the conventional scheme \cite{Jiang2015}, (2) the optimized scheme with \eqref{eq:exp_design}, and (3) the optimized scheme with \eqref{eq:exp_design2}. In Fig.~\ref{fig:conv_acq}, we show the acquisition parameters from the conventional scheme with $N = 1000$, as well as the resulting magnetization evolutions for the two regions-of-interest (ROIs), respectively, in the white matter tissue and a gray matter tissue (as marked in Fig.~\ref{fig:ground_truth} (a)).
\begin{figure*}[!thb]
\centering
{\includegraphics[width=0.98\textwidth]{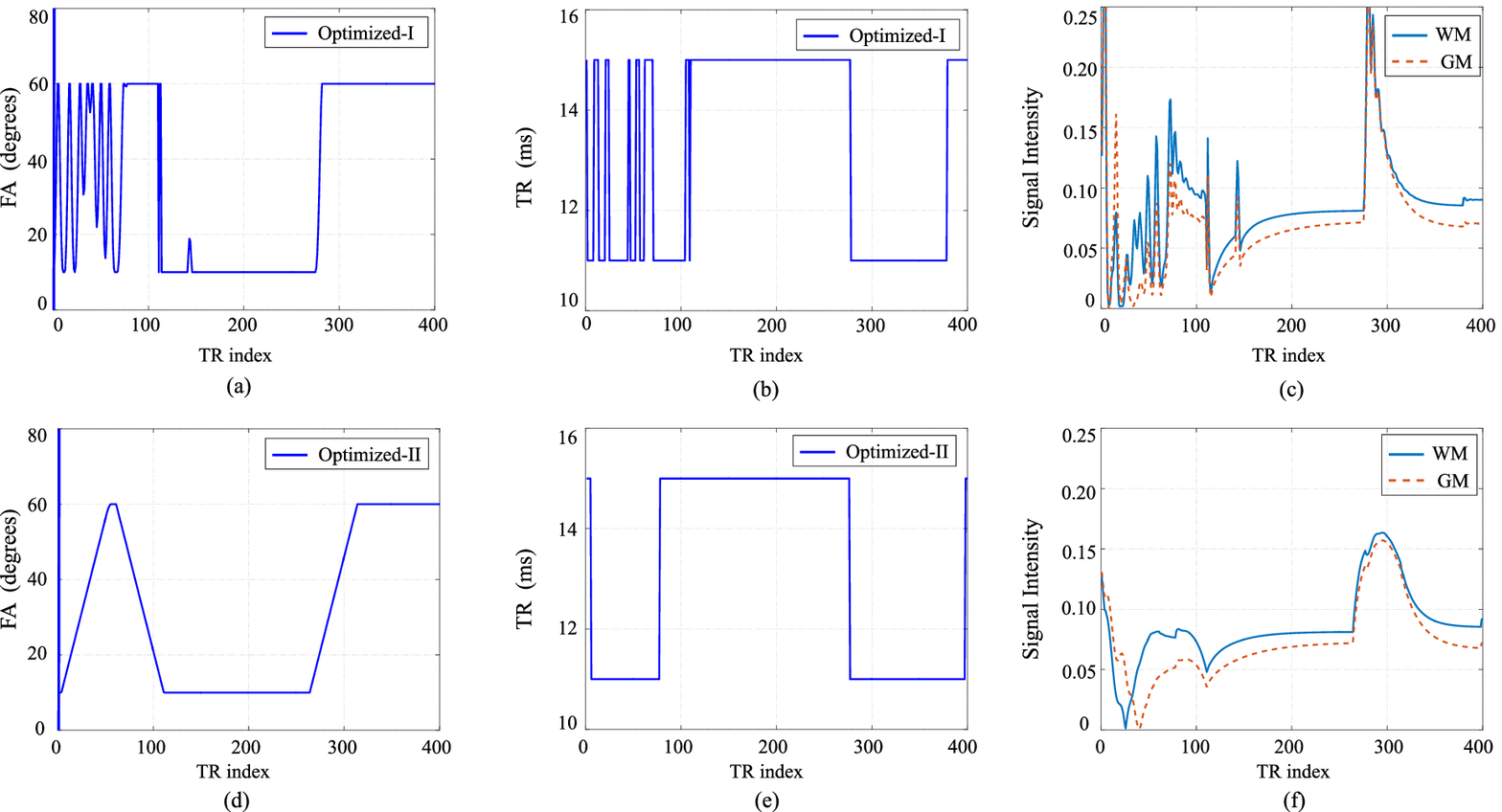}}
\caption{Two optimized acquisition schemes with $N = 400$, and the resulting magnetization evolutions for the white matter and the gray matter ROIs marked in Fig.~\ref{fig:ground_truth} (a). (a) Flip angle train in Optimized-I. (b) Repetition time train in Optimized-I. (c) Magnetization evolutions from Optimized-I. (d) Flip angle train in Optimized-II. (e) Repetition time train in Optimized-II. (f) Magnetization evolutions from Optimized-II. Note that the first RF pulses are $180^{\circ}$ for both Optimized-I and Optimized-II, which exceed the scale of the vertical axes in (a) and (d).}
\label{fig:optim_acq_Nfr400}
\end{figure*}

\begin{figure}[!tb]
\centering
{\includegraphics{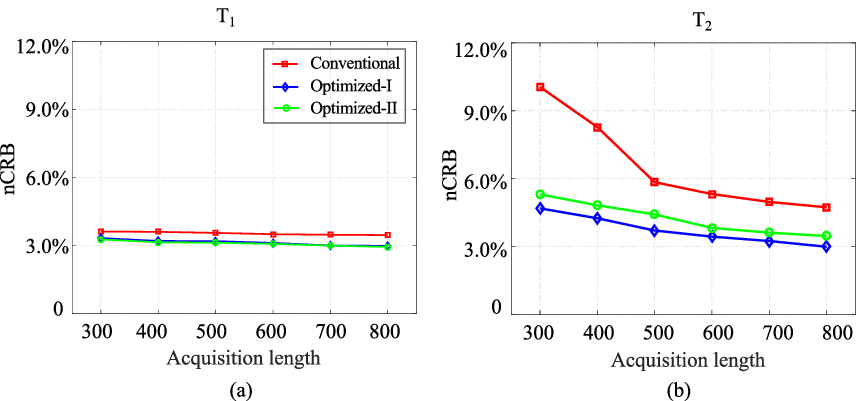}}
\caption{Normalized CRB versus acquisition length $N$ for the ROI in the white matter tissue. (a) Normalized CRBs for $T_1$. (b) Normalized CRBs for $T_2$.}
\label{fig:Tissue1_nCRB_vs_Nfr}
\end{figure}

We generated $\mathbf{k}$-space data from the contrast-weighted images using the non-uniform Fourier transform \cite{fessler2003}. In the simulations, we considered two setups of MR fingerprinting experiments:
\begin{itemize}
\item Fully-sampled experiment: we acquired fully-sampled Cartesian $\mathbf{k}$-space data for each time point (or TR index), as in \cite{Gao2015}.
\item Highly-undersampled experiment: we acquired highly-undersampled spiral $\mathbf{k}$-space data at each time point, as in \cite{Jiang2015}. For this, we used the same spiral trajectory as \cite{Jiang2015}, and acquired only one spiral interleaf for each time point, whereas a full set of spiral trajectory consists of 48 interleaves.
\end{itemize}

We added complex white Gaussian noise to the measured $\mathbf{k}$-space data according to the pre-specified noise level $\sigma^2$. Here we define the signal-to-noise ratio as $\mathrm{SNR} = 20\log_{10}{(s/\sigma)}$, where $s$ denotes the average value of $M_0$ in a region of white matter. This definition measures the SNR in decibels (dB).

We performed the ML reconstruction \cite{zhao2015, zhao2016} for the above experiments. Note that for the fully-sampled Cartesian experiments, the ML approach is equivalent to the direct Fourier reconstruction, followed by the dictionary-based pattern matching \cite{Haldar2007}. For the highly-undersampled experiments, we solved the reconstruction problem with the algorithm in \cite{zhao2015, zhao2016}, which we initialized with the gridding reconstruction. Here the dictionary used in the ML reconstruction was constructed based on the following parameter discretization scheme: we set the $T_1$ value in the range $\left[20, 3000\right]\mathrm{ms}$, in which we used an increment of 10~ms for $\left[20, 1500\right]\mathrm{ms}$ and an increment of 30~ms for $\left[1501, 3000\right]\mathrm{ms}$; we set the $T_2$ value in the range $\left[30, 500\right]\mathrm{ms}$, in which we used an increment of 1~ms for $\left[30, 200\right]\mathrm{ms}$ and an increment of 5~ms for $\left[201, 500\right]\mathrm{ms}$.

To assess the reconstruction accuracy, we used the following two metrics: (a) overall error, i.e., $\|\mathbf{I} - \hat{\mathbf{I}}\|_2/\|\mathbf{I}\|_2$, where $\mathbf{I}$ and $\hat{\mathbf{I}}$ respectively denote the true parameter map and reconstructed parameter map, and (b) voxelwise relative error, i.e., $|\mathbf{I}_{v} - \hat{\mathbf{I}}_v|/|\mathbf{I}_{v}|$, where $\mathbf{I}_{v}$ and $\hat{\mathbf{I}}_{v}$ respectively denote the values of $\mathbf{I}$ and $\hat{\mathbf{I}}$ at the $v$th voxel.

\subsubsection{Implementation of \eqref{eq:exp_design} and \eqref{eq:exp_design2}}
Here we describe the detailed implementation of the proposed approach for the above application example. For convenience, we refer to the optimized schemes with $\eqref{eq:exp_design}$ and $\eqref{eq:exp_design2}$ as Optimized-I and Optimized-II, respectively. In this work, we assumed that $T_1$ and $T_2$ were of primary interest, and chose three representative tissues, i.e., $\boldsymbol{\uptheta}^{(1)} = \left[700~ \mathrm{ms}, ~60~\mathrm{ms}, ~0.6\right]$, $\boldsymbol{\uptheta}^{(2)} = \left[850~\mathrm{ms}, ~50\mathrm{ms}, ~0.6\right]$, and $\boldsymbol{\uptheta}^{(3)} = \left[1100~\mathrm{ms}, ~102\mathrm{ms}, ~0.6\right]$, for $\eqref{eq:exp_design}$ and $\eqref{eq:exp_design2}$.\footnote{The $T_1$ and $T_2$ values of these tissues were arbitrarily chosen from the range of tissue parameter values that are relevant to neuroimaging. To avoid the inverse crime, we did not take tissue parameter values directly from the brain phantom.} Moreover, we manually chose the weighting matrix $\mathbf{W} = \mathrm{diag}([2.0\times10^{-5}, 5.0\times10^{-4}, 3.0\times10^{1}])$ for both \eqref{eq:exp_design} and \eqref{eq:exp_design2} to ensure the good performance of the optimized experiments.
\begin{figure*}[!th]
\centering
{\includegraphics{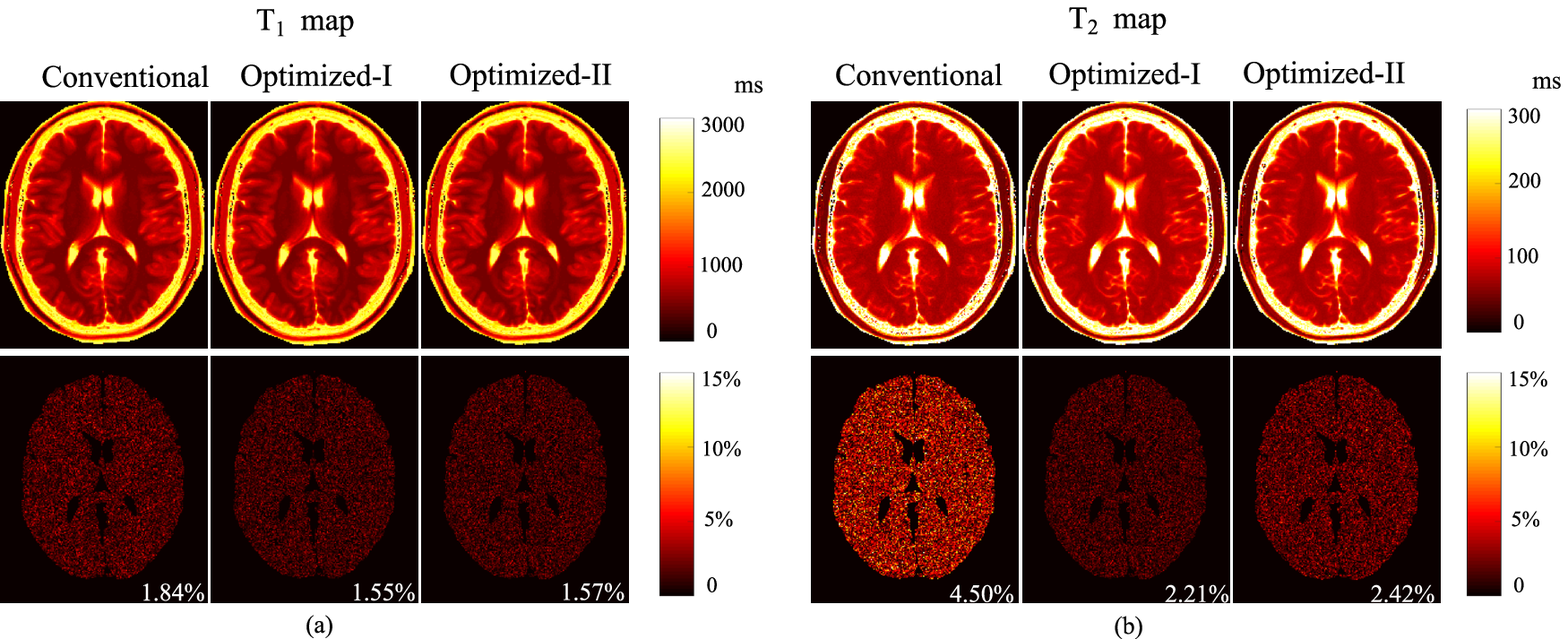}}
\caption{Reconstructed parameter maps from the fully-sampled MR fingerprinting experiments ($N = 400$ and $\mathrm{SNR} = 33~\mathrm{dB}$), using the acquisition parameters from the conventional scheme, Optimized-I, and Optimized-II. (a) Reconstructed $T_1$ maps and associated relative error maps. (b) Reconstructed $T_2$ maps and associated relative error maps. Note that the overall error is labeled at the lower right corner of each error map, and the regions associated with the background, skull, scalp, and CSF were set to be zero.}
\label{fig:sim_recon_Nfr400}
\end{figure*}

\begin{figure*}[!th]
\centering
{\includegraphics[width=0.98\textwidth]{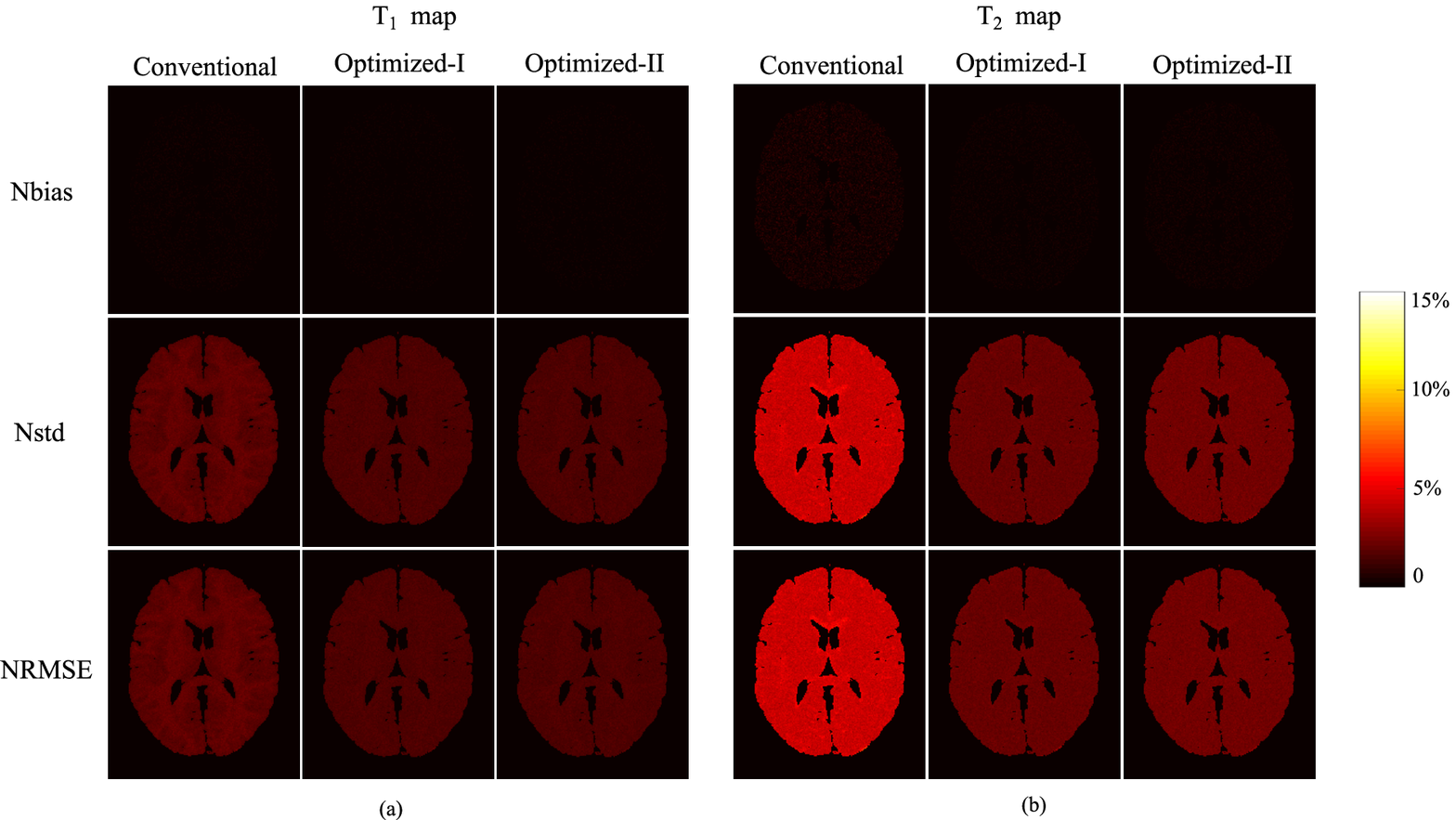}}
\caption{Bias-variance analysis of the reconstructed parameter maps from the fully-sampled MR fingerprinting experiments ($N = 400$ and $\mathrm{SNR} = 33~\mathrm{dB}$), using the acquisition parameters from the conventional scheme, Optimized-I, and Optimized-II. (a) Normalized bias, standard deviation, and root-mean-square error for (a) $T_1$ maps and (b) $T_2$ maps. The regions associated with the background, skull, scalp, and CSF were set to be zero.}
\label{fig:bias_variance_full}
\end{figure*}
Further we specify the constraints for the acquisition parameters in $\eqref{eq:exp_design}$ and $\eqref{eq:exp_design2}$. Specifically, we set the maximum flip angle as
\begin{equation}
\alpha_n^{\text{max}} = \begin{cases}
180^{\circ}, & \text{if $n = 1$},\\
60^{\circ}, & \text{if $2\leq n \leq N$}.
\end{cases} \nonumber
\end{equation}
Note that this allows an $180^{\circ}$ inversion pulse imposed at the beginning of imaging experiments, which is often advantageous for the $T_1$ estimation \cite{zhao2016b}. We set the minimum flip angle as $\alpha_n^{\text{min}} = 10^{\circ}$ for $1\leq n \leq N$. We respectively set the maximum and minimum repetition times as $TR_n^{\text{max}} = 15~\mathrm{ms}$ and $TR_n^{\text{min}} = 11~\mathrm{ms}$ for $1\leq n \leq N$. Here, it is worth mentioning that the above constraints roughly match the range of the acquisition parameters in the conventional scheme. For $\eqref{eq:exp_design2}$, we have the additional constraints on the flip angle variations, which were set as
\begin{equation}
\Delta\alpha_n^{\text{max}} = \begin{cases}
+ \infty, & \text{if $n = 1$},\\
1^{\circ}, & \text{if $2\leq n \leq N-1$}.
\end{cases} \nonumber
\end{equation}
Note that the above constraint does not restrict the flip angle variation between the inversion pulse and the second RF pulse.
\begin{figure}[!th]
\centering
{\includegraphics{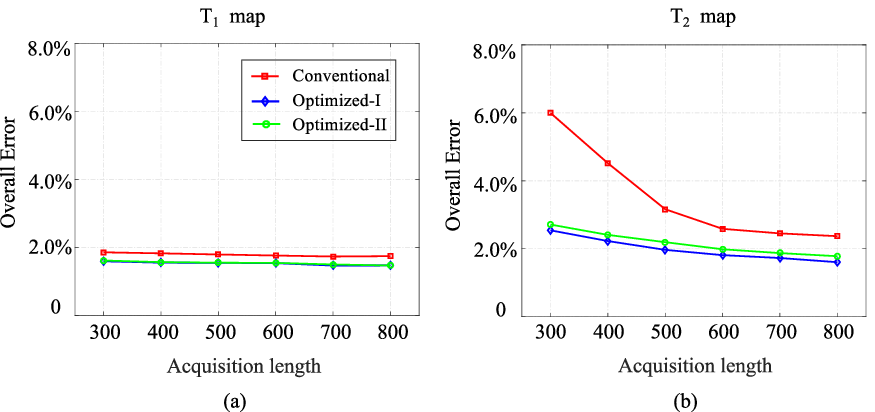}}
\caption{Overall error versus the acquisition length $N$ for the fully-sampled MR fingerprinting experiments with $\mathrm{SNR} = 33~\mathrm{dB}$. (a) Overall error of $T_1$ map. (b) Overall error of $T_2$ map. Note that the overall error is calculated with respect to the whole brain, excluding the skull, scalp, and CSF.}
\label{fig:sim_nrmse_Nacq_full}
\end{figure}

\begin{figure}[!th]
\centering
{\includegraphics{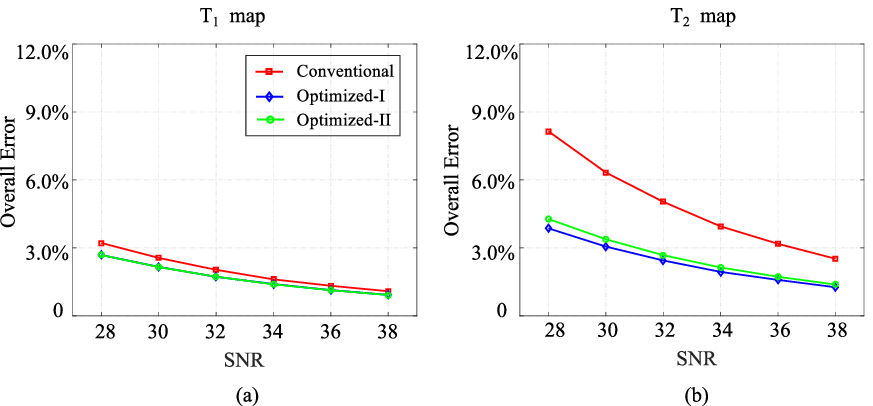}}
\caption{Overall error versus the SNR level for the fully-sampled MR fingerprinting experiments with $N = 400$. (a) Overall error of $T_1$ map. (b) Overall error of $T_2$ map. Note that the overall error is calculated with respect to the whole brain, excluding the skull, scalp, and CSF.}
\label{fig:sim_nrmse_SNR_full}
\end{figure}

We applied the SQP algorithm to solve the optimization problems associated with \eqref{eq:exp_design} and \eqref{eq:exp_design2}, and initialized the algorithm with the acquisition parameters $\left\{\alpha_n, TR_n\right\}_{n = 1}^N$ from the conventional scheme. We terminated the algorithm, when the change of the solution was less than the pre-specified tolerance (i.e., $\epsilon = 1e^{-4}$) or the maximum iteration (i.e., $J_{\mathrm{max}} = 5 \times 10^{4}$) was reached. The runtime of the algorithm depends on the length of acquisition. For example, with $N = 400$, solving \eqref{eq:exp_design} and \eqref{eq:exp_design2} respectively took about 290 min and 140 min on a Linux workstation with 24 Intel Xeon E5-2643, 3.40 GHz processors and 128 GB RAM running Matlab R2015b.

We optimized the acquisition parameters independently for several choices of $N$ (i.e., $N$ = 300, 400, 500, 600, 700, and 800). As an example, Fig.~\ref{fig:optim_acq_Nfr400} shows the optimized acquisition parameters with $N = 400$ from Optimized-I and Optimized-II. As can be seen, the optimized acquisition parameters appear to be highly structured, which are remarkably different from the acquisition parameters from the conventional scheme. In particular, the optimized repetition times turn out to be binary (i.e., switching between $TR_n^{\text{max}}$ and $TR_n^{\text{min}}$). Besides $N = 400$, we also had similar observations for all other acquisition lengths. We show one more example, i.e., the optimized schemes with $N = 600$, in the Supplementary Material. In Section~\ref{sec:discussion}, we will discuss this interesting observation.

Fig.~\ref{fig:optim_acq_Nfr400} also shows the resulting magnetization evolutions from Optimized-I and Optimized-II. As can be seen, the magnetization evolutions from Optimized-I exhibit significant oscillation due to the dramatic change of acquisition parameters (within the first 100 time points). In contrast, by enforcing the additional constraints on the flip angle variations in Optimized-II, the oscillation behavior has been significantly suppressed, and the resulting magnetization evolutions become much smoother. Later we will demonstrate that this is advantageous for image reconstruction from highly-undersampled data.

\subsubsection{Evaluation of CRB}
We evaluated the CRB for the conventional scheme, Optimized-I, and Optimized-II. Note that the CRB is a lower bound on the variance (or equivalently, the mean-square error) of an unbiased estimator, which characterizes the SNR property of an imaging experiment. Specifically, we calculated the CRB associated with the three sets of acquisition parameters over different acquisition lengths. For all the experiments, we set $\mathrm{SNR} = ~33 ~\mathrm{dB}$. In order to show the CRB for both $T_1$ and $T_2$ at the same scale, we used the normalized CRB defined as $\mathrm{nCRB}_i = \sqrt{\mathrm{CRB}(\boldsymbol{\uptheta}_i)}/\boldsymbol{\uptheta}_i$.

Fig.~\ref{fig:Tissue1_nCRB_vs_Nfr} shows the normalized CRB versus the acquisition length for the white matter ROI. As expected, for all three schemes, the nCRBs of $T_1$ and $T_2$ reduce as the acquisition length becomes longer. For $T_1$, the nCRB rapidly approaches its asymptotic limit with short acquisitions for all three schemes. However, for $T_2$, the conventional scheme needs much longer acquisition to attain a good nCRB. This is consistent with the previous observations in \cite{zhao2015, zhao2016, Eric2016, Cao2017, zhao2017}, and indicates that from an estimation-theoretic perspective, there is significant room for improvement over the conventional scheme. In contrast, the two optimized schemes significantly improve the nCRB of $T_2$, especially with short acquisition lengths. For example, the optimized experiments reduce the nCRB of $T_2$ by about a factor of two for $N = 400$.

Also note that the nCRB for Optimized-I is better than for Optimized-II over all the acquisition lengths. This is as expected, since, with a smaller set of constraints, Optimized-I searches over a larger feasible space of acquisition parameters. Nonetheless, the difference between the two optimized schemes is quite small. In the Supplementary Material, we also show the CRB evaluation with respect to the gray matter ROI (marked in Fig.~\ref{fig:ground_truth} (a)), from which we had similar observations.

\subsubsection{Evaluation of fully-sampled experiments}
We evaluated the fully-sampled MR fingerprinting experiments described in Section~\ref{sec:results}.A. Note that this scenario exactly corresponds to the data model in \eqref{eq:data_model}, with which we calculated the CRB and performed experiment design. Specifically, we simulated the experiments at $N = 400$ and $\mathrm{SNR} = 33$ dB, using the conventional scheme, Optimized-I, and Optimized-II. Fig.~\ref{fig:sim_recon_Nfr400} shows the reconstructed $T_1$ and $T_2$ maps. As can be seen, the two optimized schemes improve the accuracy of both $T_1$ and $T_2$, although the improvement is more significant for $T_2$. Moreover, the performance of Optimized-I is slightly better than that of Optimized-II, consistent with the CRB prediction shown before.
\begin{figure*}[!th]
\centering
{\includegraphics{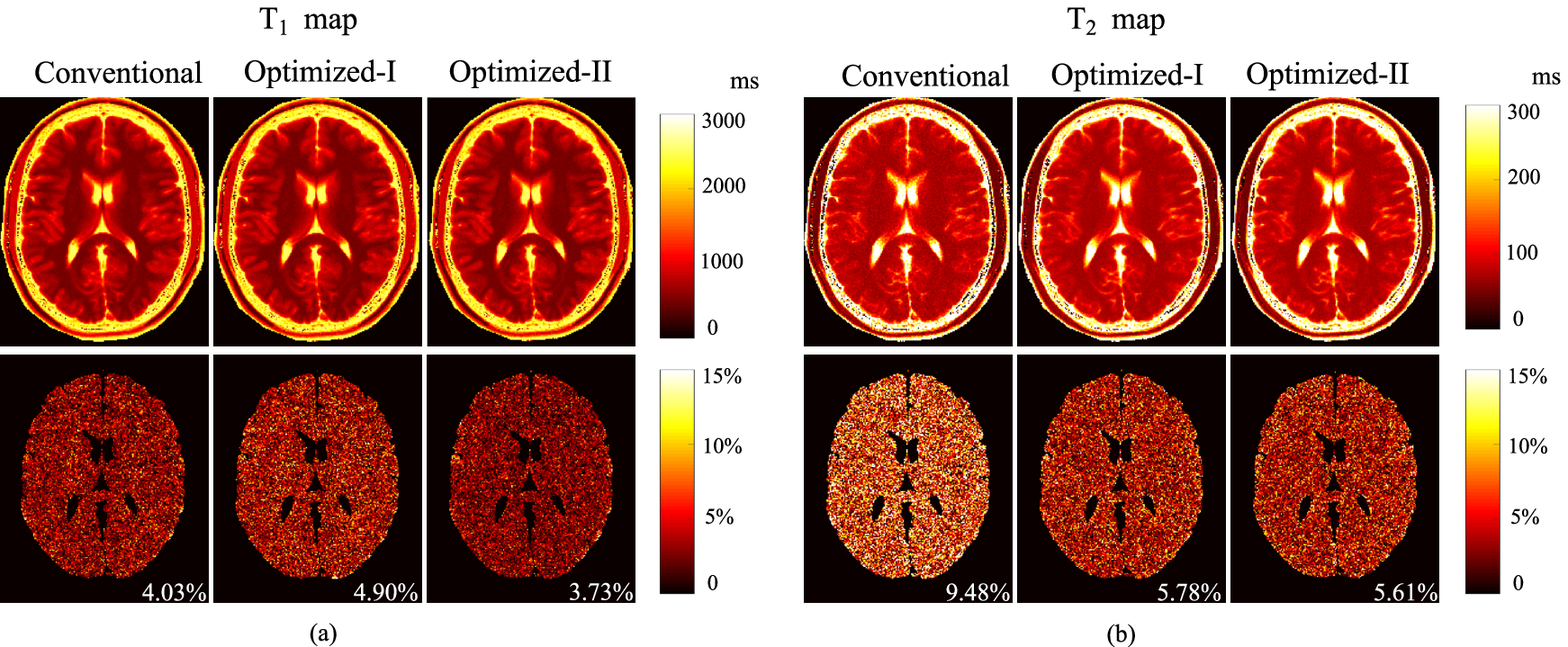}}
\caption{Reconstructed parameter maps from the highly-undersampled MR fingerprinting experiments ($N = 400$ and $\mathrm{SNR} = 33~\mathrm{dB}$), using the acquisition parameters from the conventional scheme, Optimized-I, and Optimized-II. (a) Reconstructed $T_1$ maps and associated relative error maps. (b) Reconstructed $T_2$ maps and associated relative error maps. Note that the overall error is labeled at the lower right corner of each error map, and the regions associated with the background, skull, scalp, and CSF were set to be zero.}
\label{fig:sim_recon_Nfr400_sparse}
\end{figure*}

We further investigated the bias-variance property of the reconstructed parameter maps. Specifically, we performed Monte Carlo (MC) simulations with 100 trials to calculate the bias, standard deviation, and root-mean-square error of the reconstructed parameter maps. For convenience, we normalized these quantities as follows:
(1) normalized bias:
\begin{equation}
\text{Nbias}_v = \hat{\mathbb{E}}\left[\left|\mathbf{I}_v - \hat{\mathbf{I}}_v\right|\right]/\mathbf{I}_v,
\end{equation}
(2) normalized standard deviation:
\begin{equation}
\label{eq:nstd}
\text{Nstd}_v = \sqrt{\hat{\mathbb{E}}\left[\left|\hat{\mathbf{I}}_v - \hat{\text{E}}(\hat{\mathbf{I}}_v)\right|^2\right]}/\mathbf{I}_v,
\end{equation}
and (3) normalized root-mean-square error:
\begin{equation}
\text{NRMSE}_v = \sqrt{\hat{\mathbb{E}}\left[\left|\mathbf{I}_v - \hat{\mathbf{I}}_v\right|^2\right]}/\mathbf{I}_v,
\end{equation}
where $\hat{\mathbb{E}}(\cdot)$ denotes the empirical mean evaluated for the MC simulations, and $\mathbf{I}_v$ and $\hat{\mathbf{I}}_v$ respectively denote the $v$th voxel from the true parameter map and reconstructed parameter map. Note that
\begin{equation}
\text{NRMSE}_v = \sqrt{\text{Nbias}_v^2 + \text{Nstd}^2_v}. \nonumber
\end{equation}
Fig.~\ref{fig:bias_variance_full} shows the normalized bias, standard deviation, and root-mean-square error maps for the reconstructed $T_1$ and $T_2$ maps. As can be seen, Optimized-I and Optimized-II reduce the normalized standard deviations for both $T_1$ and $T_2$, compared to the conventional scheme. Consistent with the CRB prediction and the results shown in Fig.~\ref{fig:sim_recon_Nfr400}, the improvement for $T_2$ is more substantial than for $T_1$. Moreover, for all three acquisition schemes, the normalized standard deviation is much larger than the normalized bias, and the normalized root-mean-square error is dominated by the normalized standard deviation. The above behavior can be expected, given that the ML reconstruction is asymptotically unbiased \cite{kay1993}.

Moreover, we evaluated the fully-sampled experiments with different acquisition lengths, with $N$ ranging from 300 to 800. Here we set $\mathrm{SNR} = 33~$dB for the experiments. Fig.~\ref{fig:sim_nrmse_Nacq_full} shows the overall errors of $T_1$ and $T_2$ versus the acquisition length. Clearly, the two optimized acquisition schemes outperform the conventional scheme over all the acquisition lengths. As one more example, we show the reconstruction results for $N = 600$ in the Supplementary Material.

Finally, we evaluated the fully-sampled experiments with different SNR levels, ranging from 28 dB to 38 dB. Here we set at $N = 400$ for all the experiments. Fig.~\ref{fig:sim_nrmse_SNR_full} shows the overall errors of $T_1$ and $T_2$ versus the SNR. As can be seen, the optimized acquisition schemes outperform the conventional scheme over all the SNR levels.
\begin{figure*}[!th]
\centering
{\includegraphics[width=0.98\textwidth]{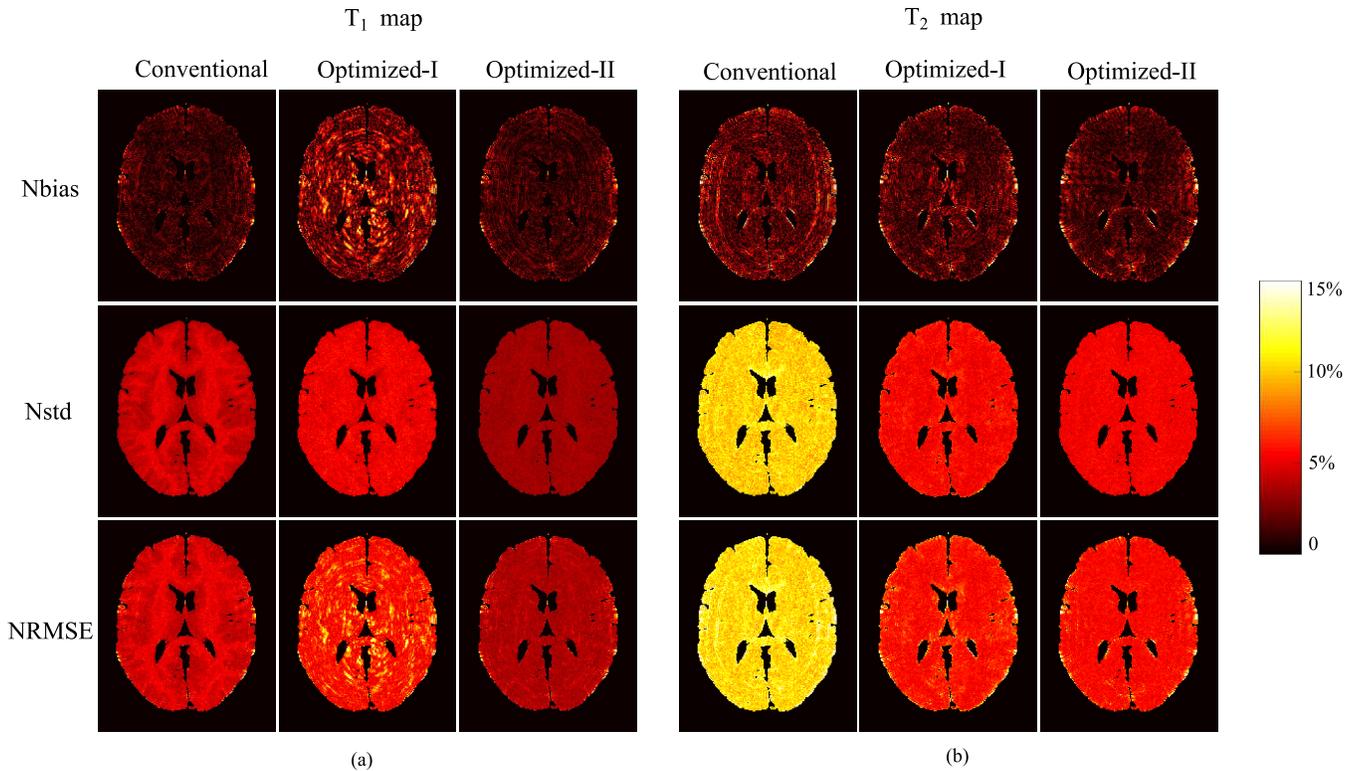}}
\caption{Bias-variance analysis of the reconstructed parameter maps from the highly-undersampled MR fingerprinting experiments ($N = 400$ and $\mathrm{SNR} = 33~\mathrm{dB}$), using the acquisition parameters from the conventional scheme, Optimized-I, and Optimized-II. (a) Normalized bias, standard deviation, and root-mean-square error for (a) $T_1$ maps and (b) $T_2$ maps. The regions associated with the background, skull, scalp, and CSF were set to be zero.}
\label{fig:bias_variance_sparse}
\end{figure*}
\subsubsection{Evaluation of highly-undersampled experiments}
We repeated the same evaluations but applied to the highly-undersampled case. Note that this more closely matches the way that MR fingerprinting is applied in practice. Fig.~\ref{fig:sim_recon_Nfr400_sparse} shows the reconstructed $T_1$ and $T_2$ maps from the highly-undersampled experiments at $N = 400$ and $\mathrm{SNR} = 33$ dB, using the conventional scheme, and the two optimized schemes. As can be seen, Optimized-I improves the accuracy of the $T_2$ map over the conventional scheme, but at the expense of degrading the accuracy of the $T_1$ map. In contrast, Optimized-II provides better accuracy for both $T_1$ and $T_2$ maps, which is highly desirable. Note that the ML reconstruction involves solving a nonlinear and nonconvex optimization problem, for which a good initialization is often required. By enforcing the additional constraint on the flip angle variations, Optimized-II results in much smoother magnetization evolutions (as shown in Fig.~\ref{fig:optim_acq_Nfr400}). With the highly-undersampled data, this often leads to better pattern matching results for the conventional reconstruction, which in turn provides an improved initialization for the ML reconstruction.
\begin{figure}[!t]
\centering
{\includegraphics{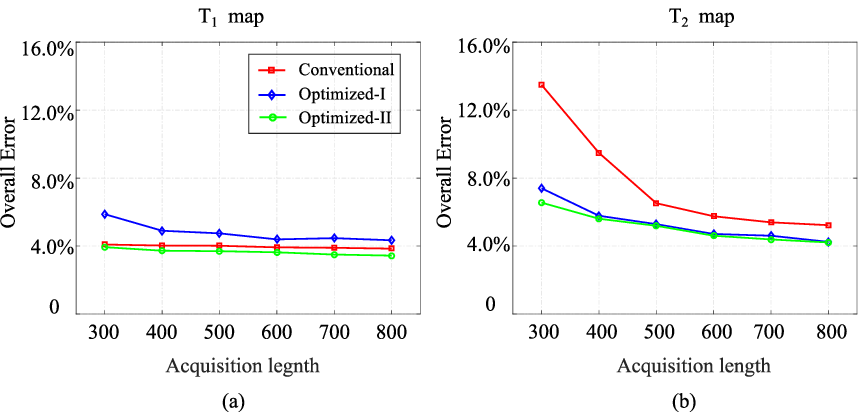}}
\caption{Overall error versus the acquisition length $N$ for the highly-undersampled MR fingerprinting experiments with $\mathrm{SNR} = 33~\mathrm{dB}$. (a) Overall error of $T_1$ map. (b) Overall error of $T_2$ map. Note that the overall error is calculated with respect to the whole brain, excluding the skull, scalp, and CSF.}
\label{fig:sim_nrmse_Nacq_sparse}
\end{figure}

\begin{figure}[!t]
\centering
{\includegraphics{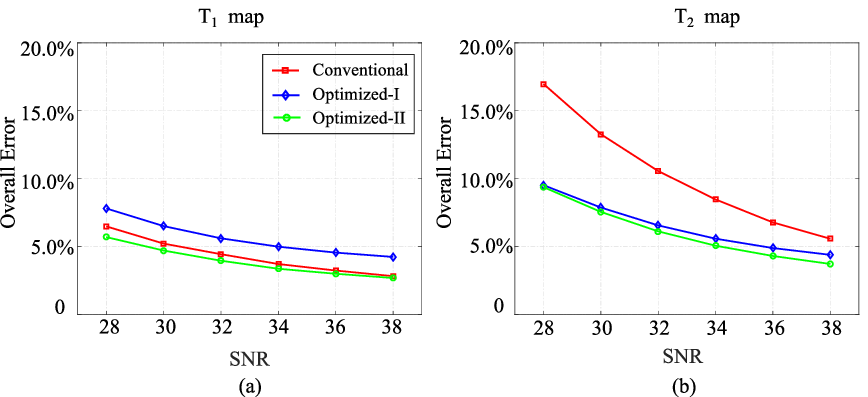}}
\caption{Overall error versus the SNR level for the highly-undersampled MR fingerprinting experiments with $N = 400$. (a) Overall error of $T_1$ map. (b) Overall error of $T_2$ map. Note that the overall error is calculated with respect to the whole brain, excluding the skull, scalp, and CSF.}
\label{fig:sim_nrmse_SNR_sparse}
\end{figure}
Fig.~\ref{fig:bias_variance_sparse} shows the normalized bias, standard deviation, and root-mean-square error maps for the reconstructed $T_1$ and $T_2$ maps from the MC simulations (with 100 trials). Clearly, Optimized-II reduces the normalized standard deviation and root-mean-square error for both $T_1$ and $T_2$ maps, compared to the conventional scheme. Moreover, with smooth magnetization evolutions, Optimized-II reduces the bias compared to Optimized-I. This further illustrates the merit of introducing the constraint on the flip angle variations for the highly-undersampled experiments.

Fig.~\ref{fig:sim_nrmse_Nacq_sparse} shows the overall errors of $T_1$ and $T_2$ versus the acquisition length, for the highly-undersampled experiments with $\mathrm{SNR} = 33~$dB. Clearly, Optimized-II outperforms the conventional scheme and Optimized-I for all the acquisition lengths. As a further illustration, we show the reconstruction results for $N = 600$ in the Supplementary Material.

Fig.~\ref{fig:sim_nrmse_SNR_sparse} shows the overall errors of $T_1$ and $T_2$ versus the SNR, for the highly-undersampled experiments with $N = 400$. As can be seen, Optimized-II provides better accuracy than the conventional scheme and Optimized-I over all the SNR levels.
\begin{figure*}[!th]
\centering
{\includegraphics[width=0.92\textwidth]{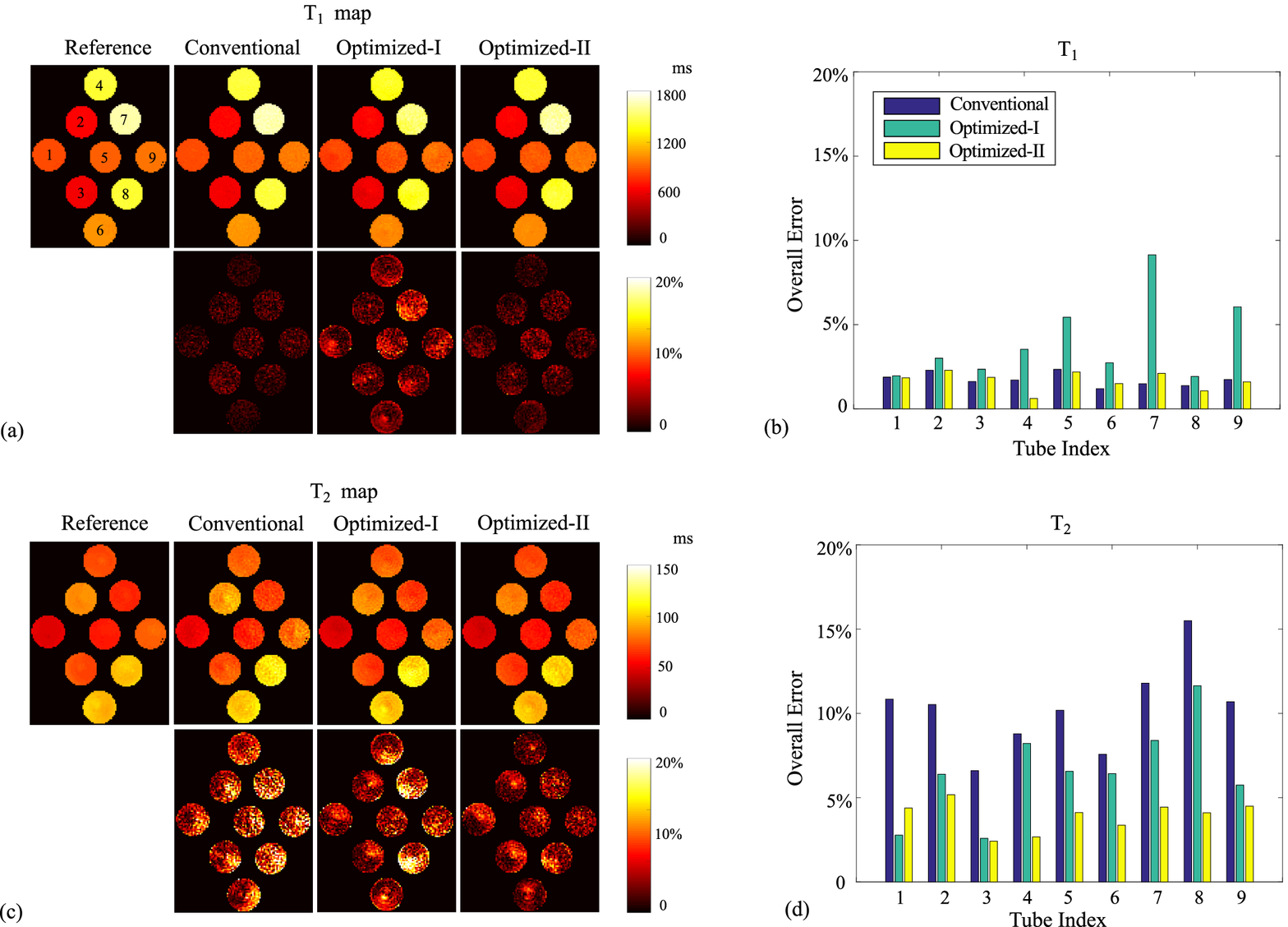}}
\caption{Reconstructed parameter maps for the phantom experiments ($N = 400$) using the acquisition parameters from the conventional scheme, Optimized-I, and Optimized-II. (a) Reconstructed $T_1$ map and associated relative error map. (b) Overall error of $T_1$ reconstruction over each tube. (c) Reconstructed $T_2$ map and associated relative error map. (d) Overall error of $T_2$ over each tube. Note that the highly-undersampled MR fingerprinting experiments using the conventional scheme, Optimized-I, and Optimized-II respectively took 5.28 sec, 5.24 sec, and 5.22 sec, whereas the fully-sampled MR fingerprinting experiment for the reference data took 18.3 min.}
\label{fig:phantom_exp_results}
\end{figure*}
\subsection{Phantom Experiments}
\label{sec:invivo_results}
We evaluated the proposed approach with phantom experiments. Here we focus on the scenario of highly-undersampled MR fingerprinting experiments, which is of the most practical interest for quantitative MR imaging. Specifically, we created a physical phantom that consists of 9 plastic tubes, each one filled with a solution of Gadolinium and Agar at different concentrations. This created different combinations of $T_1$ and $T_2$ values that are relevant to the neuroimaging application \cite{Kengo2013}. We carried out the experiments on a 3T Siemens Tim Trio scanner (Siemens Medical Solutions, Erlangen, Germany) equipped with a 32-channel head array coil. The relevant imaging parameters include: FOV = $300\times 300~\mathrm{mm}^2$, matrix size = $256\times 256$, and slice thickness = $5~\mathrm{mm}$.

\subsubsection{General evaluation}
We performed three sets of experiments with $N = 400$, respectively, using the acquisition parameters from the conventional scheme, Optimized-I, and Optimized-II. We used the same spiral trajectory and sampling pattern as in the numerical simulations. Here the acquisition times for the conventional scheme, Optimized-I, and Optimized-II were 5.28 sec, 5.24 sec, and 5.22 sec, respectively. To evaluate the performance of the above experiments, we also acquired a set of reference $T_1$ and $T_2$ maps, by performing a fully-sampled MR fingerprinting experiment\footnote{The fully-sampled experiment was performed by repeating a highly-undersampled acquisition 48 times. For each acquisition, we switched to a different spiral interleaf at every time point. Note that a short time delay was added between consecutive acquisitions to ensure that the magnetization starts at thermal equilibrium.} with $N = 1000$ using the acquisition parameters from the conventional scheme. The acquisition time for this experiment was about 18 min. Additionally, we calibrated the spiral trajectory with a specialized pulse sequence \cite{Hao2009} to avoid the potential trajectory distortion (caused by eddy currents and gradient delay). Finally, we performed an auxiliary scan with a gradient echo (GRE) sequence, from which we estimated the coil sensitivity maps. This acquisition took 1.28 sec.

We performed the ML reconstruction for the above experiments, and incorporated the slice-profile correction \cite{Dan2017} into the dictionary. Fig.~\ref{fig:phantom_exp_results} shows the reconstructed $T_1$ and $T_2$ maps, the relative error maps (evaluated with respect to the reference data), and the corresponding reconstruction errors over each tube. It is clear that Optimized-II significantly improves the accuracy of the $T_2$ map over the conventional scheme, while providing similar accuracy for the $T_1$ map. Compared to Optimized-I, Optimized-II also provides better accuracy, particularly for $T_1$ maps. This confirms the benefits of introducing the constraint on the flip angle variation for highly-undersampled MR fingerprinting experiments.

\subsubsection{Evaluation of cross-scan variance}
\begin{figure}[!th]
\centering
{\includegraphics[width=0.48\textwidth]{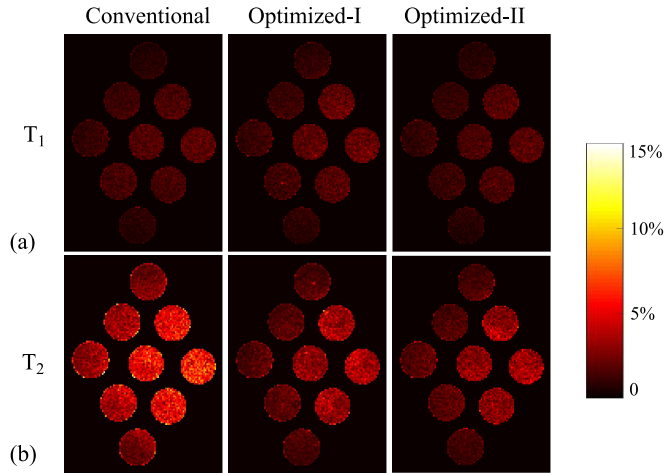}}
\caption{Normalized standard deviation maps associated with the conventional acquisition, Optimized-I, and Optimized-II, estimated from 15 independent imaging experiments. (a) Normalized standard deviation maps for $T_1$. (b) Normalized standard deviation maps for $T_2$.}
\label{fig:phantom_exp_variance}
\end{figure}

\begin{figure}[!h]
\centering
{\includegraphics[width=0.38\textwidth]{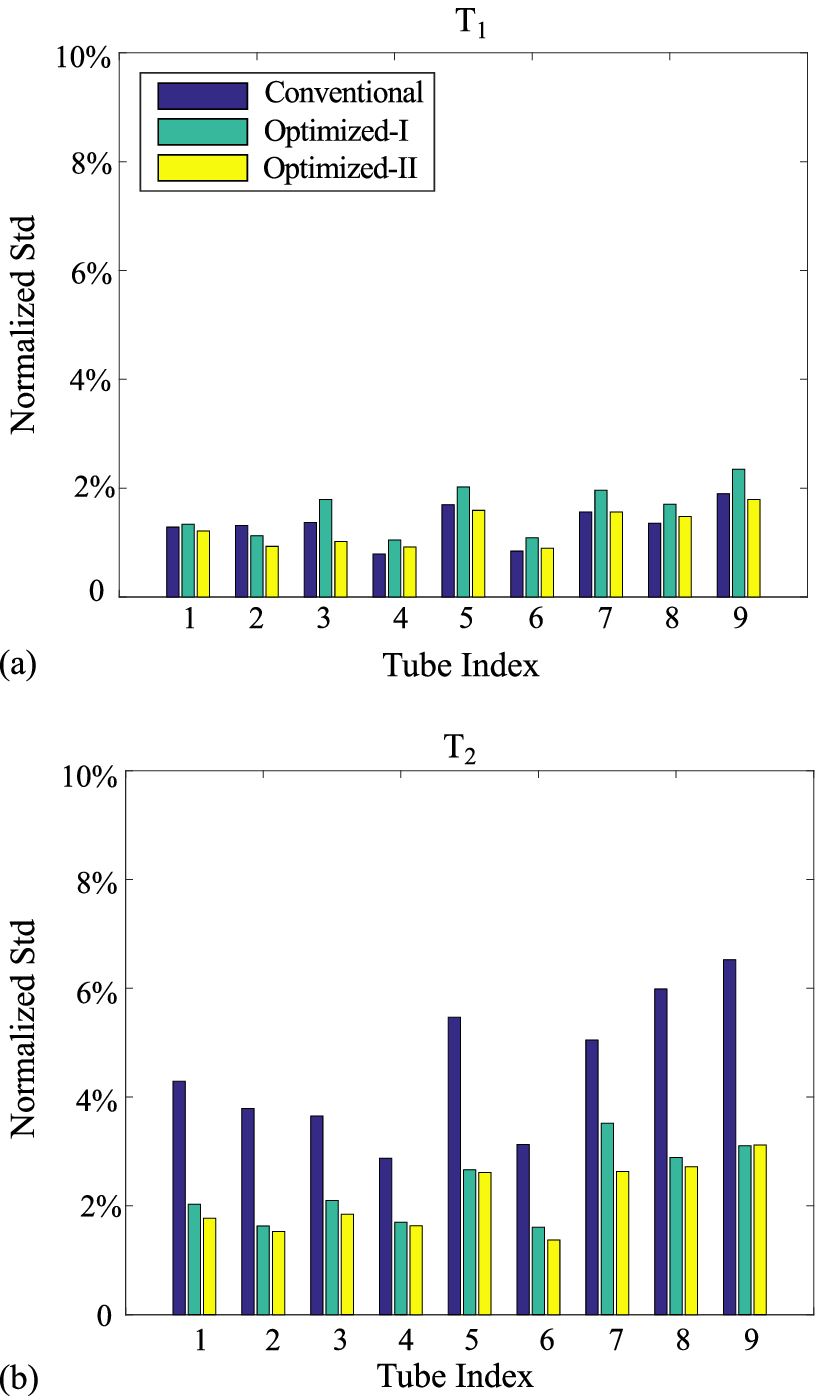}}
\caption{Normalized standard deviation averaged over each tube for (a) $T_1$ and (b) $T_2$.}
\label{fig:phantom_exp_variance_hist}
\end{figure}
Given that variance reduction is a direct benefit of the CRB based experiment design \cite{Nataraj2017}, we evaluated the variance associated with the three acquisition schemes. Specifically, we conducted each acquisition 15 times, and calculated the normalized standard deviation associated with each acquisition scheme. With the absence of the ground truth for the phantom experiments, the standard deviations were normalized with respect to the parameter maps reconstructed from the fully-sampled data with $N = 1000$. Fig.~\ref{fig:phantom_exp_variance} shows the normalized standard deviation maps for the three acquisition schemes, and Fig.~\ref{fig:phantom_exp_variance_hist} shows the normalized standard deviation averaged over each tube. It is evident that the two optimized experiments substantially reduce the standard deviation for $T_2$ maps, while providing similar standard deviation for $T_1$ maps. In particular, Optimized-II achieves a factor of two reduction in the standard deviation of $T_2$ maps for all the tubes.

\subsection{In Vivo Experiments}
We evaluated the performance of the proposed approach for in vivo experiments. We conducted the imaging experiments on a healthy volunteer with the approval from the local Institutional Review Board and the informed consent was obtained from the subject. We performed highly-undersampled MR fingerprinting experiments with the conventional scheme, Optimized-I, and Optimized-II on the same scanner with the acquisition length $N = 400$. We performed the ML reconstruction to estimate the $T_1$ and $T_2$ maps from highly-undersampled data. Moreover, we repeated each acquisition 15 times, and evaluated the sample variance associated with each acquisition scheme.

\begin{figure*}[!t]
\centering
{\includegraphics[width=0.92\textwidth]{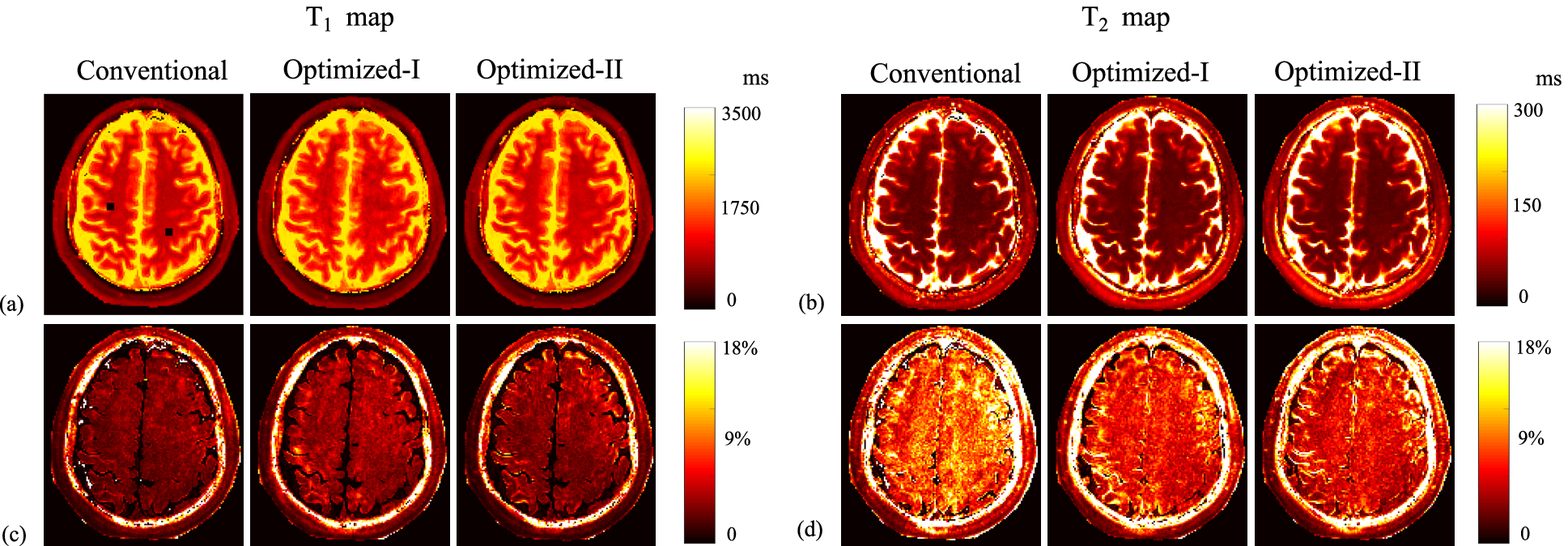}}
\caption{Reconstructed parameter maps and normalized standard deviation maps for the in vivo experiments ($N = 400$) using the acquisition parameters from the conventional scheme, Optimized-I, and Optimized-II. (a) Reconstructed $T_1$ maps. (b) Reconstructed $T_2$ maps. (c) Normalized standard deviation maps for $T_1$. (d) Normalized standard deviation maps for $T_2$.}
\label{fig:invivo_exp_results}
\end{figure*}

\begin{figure*}[!t]
\centering
{\includegraphics[width=0.75\textwidth]{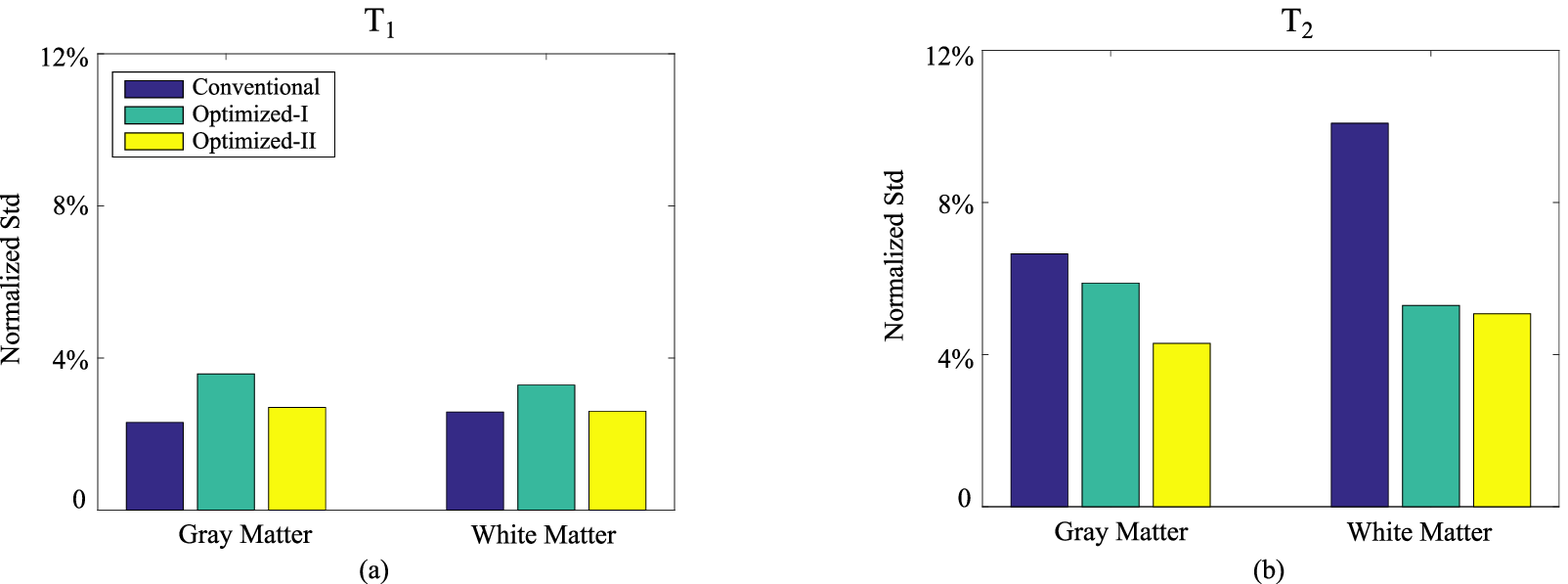}}
\caption{Normalized standard deviation averaged over the ROIs in the gray matter and white matter (marked in Fig.~\ref{fig:invivo_exp_results}) for (a) $T_1$ and (b) $T_2$.}
\label{fig:invivo_var_barplot}
\end{figure*}

Fig.~\ref{fig:invivo_exp_results} shows the reconstructed $T_1$ and $T_2$ maps from the three acquisition schemes, as well as the normalized standard deviation maps from the repetitions of the imaging experiments. Here the standard deviations were all normalized with respect to the parameter maps reconstructed from the conventional scheme with $N = 400$. Fig.~\ref{fig:invivo_var_barplot} shows the normalized standard deviations averaged over the two representative ROIs respectively in the gray matter and white matter. Additional results comparing the tissue parameter standard deviations for the whole brain can be found in the Supplementary Material.

First, note that the two optimized acquisition schemes provide the $T_1$ and $T_2$ maps in a similar range as the conventional scheme, confirming the feasibility of the proposed experiment designs. Second, the two optimized acquisition schemes improve the standard deviation for $T_2$, while providing similar standard deviation for $T_1$. The above results are consistent with those for the numerical simulations and phantom experiments. Lastly, comparing with Optimized-I, Optimized-II reduces the artifacts associated with the $T_1$ reconstruction in the white matter, and also enables better variance for both $T_1$ and $T_2$ estimation. This further illustrates the benefits of introducing the flip variation constraint in the experiment design.

\section{Discussion}
\label{sec:discussion}

It is worth discussing several related aspects as well as the potential extensions of the proposed framework. First, regarding the model of spin dynamics, we introduced the state-space model based on the isochromat summation approach \cite{Pavel1997, Malik2016}. Alternatively, we can derive the model from the expanded phase graph (EPG) formalism \cite{Matthias2015, Jiang2015}. Similar to the early work \cite{Malik2016}, we have observed that the two models are equivalent in terms of simulating magnetization evolutions (see the Supplementary Material for the numerical justification). However, the state-space model from the isochromat summation approach is conceptually simple, and easy to describe mathematically, which facilitates the subsequent CRB derivation.

Also note that our work used a simplified data model \eqref{eq:data_model} to calculate the CRB and to optimize acquisition parameters. This yields very small-scale FIMs, enabling efficient computation and storage. To tailor for highly-undersampled experiments, we further incorporated the constraint on the flip angle variation to balance the improvement of encoding efficiency and the considerations for decoding. As a generalization, it would be interesting to formulate the problem that directly optimizes highly-undersampled MR fingerprinting experiments. This could allow a joint design of acquisition parameters and $\mathbf{k}$-space trajectories (e.g., \cite{Dan2005, chao2011}). However, note that the CRB calculation in this case requires a significantly higher-dimensional formulation that models both contrast encoding and spatial encoding \cite{zhao2014}. This often results in far more expensive computation, which motivates the investigation of advanced computational algorithms (e.g., \cite{Fessler1994, Tune2012}).

For the proposed formulation, we manually selected the parameters of the weighting matrix and the constraint sets, which provided the good performance for the application example. In general, we should choose these parameters according to application-specific requirements. For example,  it may be desirable to constrain flip angles to be small, and apply multiple inversion pulses during acquisition for cardiac MR fingerprinting experiments (e.g., \cite{Hamilton2017}).

Another related aspect is that the example implementation of our proposed approach utilizes only a few representative tissues to design MR fingerprinting experiments. This is practical, since the range of tissue parameter values is often known a priori for a specific imaging application. This strategy generalizes well, as demonstrated in the numerical simulations and real experiments. In the Supplementary Material, we include additional results to further confirm the good generalization capability of this strategy. Alternatively, we can incorporate prior tissue parameters into the experiment design in other ways. For example, we can perform a Bayesian experiment design by incorporating a probability distribution of tissues parameter values \cite{Christina2016}, and perform optimal experiment design with the Bayesian CRB \cite{Vantrees2007}. Other alternatives include the min-max experiment design that minimizes the worst-case performance \cite{Nataraj2017} or average variance experiment design that minimizes the average-case performance across an interval of parameter values \cite{haldar2009b}.

Regarding the numerical optimization, it is worth reiterating that the proposed formulations in \eqref{eq:exp_design} and \eqref{eq:exp_design2} result in non-convex optimization problems, and that the solution algorithm is only guaranteed to have the local convergence. In this work, we initialized the algorithm with the acquisition parameters from conventional MR fingerprinting experiments, which demonstrated the good performance. In the Supplementary Material, we evaluated the impact of initialization on the experiment design. Simply put, we find that different initializations for \eqref{eq:exp_design} can lead to different local minima, although the cost function values of these local minima are very close. More interestingly, we find that for \eqref{eq:exp_design2}, several very distinct initializations produce the same optimized acquisition parameters. For a non-convex optimization problem, this behavior is rather remarkable, which is worth an in-depth theoretical study in the future.

By solving \eqref{eq:exp_design} and \eqref{eq:exp_design2}, we observed that the optimized acquisition parameters appeared to be highly structured. In particular, the optimized repetition times appeared to be binary. Such an interesting behavior is worth an in-depth study. Note that as shown in \cite{Chen1975, Maidens2016}, certain CRB based experiment design problem involving a dynamic system can be cast as an optimal control problem. It is possible that our empirical observations could be explained by the principles of optimal control theory \cite{Bertsekas2012, Osmolovskii2012}.\footnote{To our knowledge, this work is the first that reports Bang-Bang structure for the optimized acquisition parameters in MR fingerprinting. Although Maidens et al. formulated the experiment design problem as an optimal control problem and derived an approximate dynamic programming algorithm \cite{Maidens2016}, they only considered a highly-simplified problem setup and did not report the Bang-Bang behavior.} Although a control-theoretic characterization of \eqref{eq:exp_design} and \eqref{eq:exp_design2} is beyond the scope of this paper, it is an interesting topic to follow up on. In particular, if we know in advance that \eqref{eq:exp_design} and \eqref{eq:exp_design2} admit structured solutions, this fact could be leveraged to simplify the algorithms used to derive an optimal experiment design.

From an optimal control viewpoint, the SQP algorithm for solving \eqref{eq:exp_design} and \eqref{eq:exp_design2} targets a necessary condition of an optimal control problem (i.e., the Pontryagin's minimum principle \cite{Bertsekas2012}). As with other nonlinear programming algorithms, the local minima obtained from the SQP algorithm are generally not as good as those that would be obtained from dynamic programming \cite{Maidens2016, Bertsekas2012}, which, in contrast, deals with a sufficient condition of an optimal control problem. However, the complexity of the SQP algorithm grows polynomially with the number of the design parameters, and thus does not have the curse-of-dimensionality issue as does dynamic programming. In practice, the SQP solutions show substantial improvement over the conventional acquisition, demonstrating their practical utility.

With respect to the experimental validation, we demonstrated the key merit of the proposed framework, i.e., variance reduction in parameter estimation, especially for $T_2$ estimation. For example, we have demonstrated that the proposed framework enables about a factor of two reduction of the standard deviation for $T_2$. Such an improvement can be substantial, which could potentially correspond to a quadrupling of averaging time, or an upgrade to a higher field scanner. We expect that it will have a positive impact on the utility of MR fingerprinting.

While the CRB optimization focuses on minimizing the variance with the assumption of unbiased estimation, the real parameter estimation error will also include contributions from bias. For numerical simulations,  we demonstrated that Optimized-II provides a similar level of bias as the conventional scheme. However, for real experiments, particularly in vivo experiments, assessing the bias can be more involved. First, note that there is no an underlying ground truth for in vivo experiments\cite{Stikov2015}. Second, various model mismatches can often complicate the bias evaluation and comparison. For example, the partial volume effect often yield different magnetization evolutions under different signal excitations, which can result in significant difficulty when comparing different acquisition schemes \cite{Stikov2015, Nataraj2017}. A thorough assessment of the bias for parameter estimation still remains an open problem for MR fingerprinting, and for quantitative MR imaging in general.

This paper systematically investigated one specific implementation of our proposed framework, but many other implementations are possible and some of these alternatives may be preferable depending on the imaging context. For example, we used the weighted A-optimality as a criterion to design MR fingerprinting experiments and demonstrated its good performance. There are alternative design criteria (e.g., the D-optimality) that could be incorporated into the proposed framework, and the exploration of these criteria may be beneficial in certain applications. As another example, we can also include other acquisition parameters (e.g., RF pulse phases, echo times, etc) into the optimization. With a larger search space of acquisition parameters, we could achieve better performance, although the resulting optimization problem will be computationally more expensive. Additionally, we could optimize MR fingerprinting experiments to include other tissue parameters, such as apparent diffusion coefficients \cite{yun2014}, magnetization transfer \cite{Wang2017, Hilbert2017}, etc. Lastly, although the implementation in this paper focused on optimizing the CRB for a fixed number of TRs, there are many other design considerations (e.g., acoustic characteristics of the pulse sequence \cite{Ma2016}, SAR, total acquisition time, etc), which are important and may be worth including within the optimization framework. Exploring these possibilities may further improve the practical significance of our approach.

\section{Conclusion}
\label{sec:conclusion}
In this work, we presented a novel estimation-theoretic framework to optimize acquisition parameters of MR fingerprinting experiments. We formulated the optimal experiment design problem that maximizes the SNR efficiency of an MR fingerprinting experiment, while incorporating additional constraints that are advantageous to the reconstruction process. The optimized experiments enables substantially improved accuracy for $T_2$ maps, while providing similar or slightly better accuracy for $T_1$ maps. Remarkably, we found that the optimized acquisition parameters appear to be highly structured, rather than random/pseudo-randomly varying as used in the conventional MR fingerprinting experiments.

\section{Acknowledgement}
The authors would like to thank the anonymous reviewers for their constructive comments, which help improve this paper. B. Zhao would like to thank Dr. P. Polo for the help with creating the physical phantom, and Drs. D. Park and Y. Chang for the discussions on the correction of acquisition imperfections.

\section{Appendix}
\label{sec:appendix}
In this appendix, we derive the Jacobian matrix $\mathbf{J}_n\left(\boldsymbol{\uptheta}\right)$ in \eqref{eq:Jacobian} for the IR-FISP sequence. This provides an example to illustrate the procedure described in \eqref{eq:derivative1} and \eqref{eq:derivative2}. Recall that in the IR-FISP sequence, $\boldsymbol{\uptheta} = \left[T_1, T_2, M_0\right]^T$ and
\begin{align}
\mathbf{J}_n\left(\boldsymbol{\uptheta}\right) = \frac{\partial{\mathbf{m}[n]}}{\partial{\boldsymbol{\uptheta}}} =
\begin{bmatrix}
\frac{\partial{}}{\partial{T_1}} \mathbf{m}[n] & \frac{\partial{}}{\partial{T_2}} \mathbf{m}[n] & \frac{\partial{}}{\partial{M_0}} \mathbf{m}[n] \nonumber
\end{bmatrix}.
\end{align}
For clarity, we first summarize the state-space model for the IR-FISP sequence as follows:
\begin{multline}
\label{eq:magnetization3}
\mathbf{M}_\mathbf{r}[n]  = \mathbf{G}(\beta_{\mathbf{r}})\mathbf{R}(T_1, T_2, TR_n)\mathbf{Q}(\alpha_n, \phi_n)\mathbf{M}_\mathbf{r}[n-1] \\
+ \frac{M_0}{N_v} \mathbf{b}(T_1, TR_n),
\end{multline}
\begin{equation}
\label{eq:magnetization4}
\mathbf{m}[n] = \sum_{\mathbf{r}}\mathbf{P}\mathbf{R}(T_1, T_2, TE_n)\mathbf{Q}(\alpha_n, \phi_n)\mathbf{M}_{\mathbf{r}}[n-1],
\end{equation}
for $n = 1, \cdots, N$.
Next, we calculate the derivatives of $\mathbf{m}[n]$ with respect to $T_1$, $T_2$, and $M_0$ based on \eqref{eq:derivative1} and \eqref{eq:derivative2}.

\subsection{Derivative of $\mathbf{m}[n]$ with respect to $T_1$}
Invoking the derivative with respect to $T_1$ on both sides of \eqref{eq:magnetization4}, we have
\begin{multline}
\label{eq:T1_derivative}
\frac{\partial\mathbf{m}[n]}{\partial T_1} =
\sum_{\mathbf{r}}\left\{\mathbf{P}\frac{\partial \mathbf{R}(T_1, T_2, TE_n)}{\partial T_1}\mathbf{Q}(\alpha_n, \phi_n)\mathbf{M}_{\mathbf{r}}[n-1] \right.\\
\left.+ \mathbf{P}\mathbf{R}(T_1, T_2, TE_n)\mathbf{Q}(\alpha_n, \phi_n)\frac{\partial \mathbf{M}_{\mathbf{r}}[n-1]}{\partial T_1}\right\},
\end{multline}
where
\begin{equation}
\frac{\partial \mathbf{R}(T_1, T_2, TE_n)}{\partial T_1} = \frac{TE_n}{T_1^2}\exp{(-\frac{TE_n}{T_1})}
\begin{bmatrix}
0&0&0\\
0&0&0\\
0&0&1
\end{bmatrix}. \nonumber
\end{equation}
Noting that
\begin{equation}
\mathbf{P}\frac{\partial \mathbf{R}(T_1, T_2, TE_n)}{\partial T_1} = \mathbf{0}, \nonumber
\end{equation}
\eqref{eq:T1_derivative} can be simplified as
\begin{equation}
\label{eq:T1_derivative2}
\frac{\partial\mathbf{m}[n]}{\partial T_1} = \sum_{\mathbf{r}}\mathbf{P}\mathbf{R}(T_1, T_2, TE_n)\mathbf{Q}(\alpha_n, \phi_n)\frac{\partial \mathbf{M}_{\mathbf{r}}[n-1]}{\partial T_1}.
\end{equation}
We then take the derivative with respect to $T_1$ on both sides of \eqref{eq:magnetization3}, i.e.,
\begin{multline}
\label{eq:T1_derivative3}
\frac{\partial \mathbf{M}_{\mathbf{r}}[n]}{\partial T_1} = \mathbf{G}(\beta_{\mathbf{r}})\frac{\partial \mathbf{R}(T_1, T_2, TR_n)}{\partial T_1}\mathbf{Q}(\alpha_n, \phi_n)\mathbf{M}_{\mathbf{r}}[n-1] \\
+ \mathbf{G}(\beta_{\mathbf{r}})\mathbf{R}(T_1, T_2, TR_n)\mathbf{Q}(\alpha_n, \phi_n)\frac{\partial \mathbf{M}_{\mathbf{r}}[n-1]}{\partial T_1}\\
+ \frac{M_0}{N_v}\frac{\partial b(T_1, TR_n)}{\partial T_1},
\end{multline}
where
\begin{equation}
\frac{\partial \mathbf{R}(T_1, T_2, TR_n)}{\partial T_1} = \frac{TR_n}{T_1^2}\exp{(-\frac{TR_n}{T_1})}
\begin{bmatrix}
0&0&0\\
0&0&0\\
0&0&1
\end{bmatrix}, \nonumber
\end{equation}
and
\begin{equation}
\frac{\partial \mathbf{b}(T_1, TR_n)}{\partial T_1} = -\frac{TR_n}{T_1^2}\exp{(-\frac{TR_n}{T_1})}
\begin{bmatrix}
0\\
0 \\
1
\end{bmatrix}.\nonumber
\end{equation}
Here we can calculate $\partial\mathbf{m}[n]/\partial T_1$ by iterating \eqref{eq:T1_derivative2} and \eqref{eq:T1_derivative3} with the initial conditions $\mathbf{M}_{\mathbf{r}}[0] = \frac{M_0}{N_v}\left[0~0~1\right]^T$ and $\partial \mathbf{M}_{\mathbf{r}}[0]/\partial T_1 = \left[0~0~0\right]^T$.
\subsection{Derivative of $\mathbf{m}[n]$ with respect to $T_2$}
Invoking the derivative respect to $T_2$ on both sides of \eqref{eq:magnetization4}, we have
\begin{multline}
\label{eq:T2_derivative}
\frac{\partial\mathbf{m}[n]}{\partial T_2} = \sum_{\mathbf{r}}\left\{\mathbf{P}\frac{\partial \mathbf{R}(T_1, T_2, TE_n)}{\partial T_2}\mathbf{Q}(\alpha_n, \phi_n)\mathbf{M}_{\mathbf{r}}[n-1] \right.\\
\left. + \mathbf{P}\mathbf{R}(T_1, T_2, TE_n)\mathbf{Q}(\alpha_n, \phi_n)\frac{\partial \mathbf{M}_{\mathbf{r}}[n-1]}{\partial T_2}\right\},
\end{multline}
where
\begin{equation}
\frac{\partial \mathbf{R}(T_1, T_2, TE_n)}{\partial T_2} =\frac{TE_n}{T_2^2}\exp{(-\frac{TE_n}{T_2})}
\begin{bmatrix}
1&0&0\\
0&1&0\\
0&0&0
\end{bmatrix}. \nonumber
\end{equation}
We then take the derivative with respect to $T_2$ on both sides of \eqref{eq:magnetization3}, i.e.,
\begin{multline}
\label{eq:T2_derivative2}
\frac{\partial \mathbf{M}_{\mathbf{r}}[n]}{\partial T_2} = \mathbf{G}(\beta_{\mathbf{r}})\frac{\partial \mathbf{R}(T_1, T_2, TR_n)}{\partial T_2}\mathbf{Q}(\alpha_n, \phi_n)\mathbf{M}_{\mathbf{r}}[n-1]\\
+ \mathbf{G}(\beta_{\mathbf{r}})\mathbf{R}(T_1, T_2, TR_n)\mathbf{Q}(\alpha_n, \phi_n)\frac{\partial \mathbf{M}_{\mathbf{r}}[n-1]}{\partial T_2},
\end{multline}
where
\begin{equation}
\frac{\partial \mathbf{R}(T_1, T_2, TR_n)}{\partial T_2} = \frac{TR_n}{T_2^2}\exp{(-\frac{TR_n}{T_2})}
\begin{bmatrix}
1&0&0\\
0&1&0\\
0&0&0
\end{bmatrix}.\nonumber
\end{equation}
Here we can calculate $\partial\mathbf{m}[n]/\partial T_2$ by iterating \eqref{eq:T2_derivative} and \eqref{eq:T2_derivative2} with the initial conditions $\mathbf{M}_{\mathbf{r}}[0] = \frac{M_0}{N_v}\left[0~0~1\right]^T$ and $\partial{\mathbf{M}_{\mathbf{r}}[0]}/{\partial T_2} = \left[0~0~0\right]^T$.
\subsection{Derivative of $\mathbf{m}[n]$ with respect to $M_0$}
Invoking the derivative with respect to $M_0$ on both sides of \eqref{eq:magnetization4}, we have
\begin{equation}
\label{eq:M0_derivative}
\frac{\partial\mathbf{m}[n]}{\partial M_0} =  \sum_{\mathbf{r}}\mathbf{P}\mathbf{R}(T_1, T_2, TE_n)\mathbf{Q}(\alpha_n, \phi_n)\frac{\partial \mathbf{M}_{\mathbf{r}}[n-1]}{\partial M_0}.
\end{equation}
We then take the derivative with respect to $M_0$ on both sides of \eqref{eq:magnetization3}, i.e.,
\begin{multline}
\label{eq:M0_derivative2}
\frac{\partial \mathbf{M}[n]}{\partial M_0} = \mathbf{G}(\beta_{\mathbf{r}})\mathbf{R}(T_1, T_2, TR_n)\mathbf{Q}(\alpha_n, \phi_n)\frac{\partial \mathbf{M}_{\mathbf{r}}[n-1]}{\partial M_0}
\\
+ \frac{1}{N_v}\mathbf{b}(T_1, TR_n).
\end{multline}
Here we can calculate $\partial\mathbf{m}[n]/\partial M_0$ by iterating \eqref{eq:M0_derivative} and \eqref{eq:M0_derivative2} with the initial condition $\partial\mathbf{M}_{\mathbf{r}}[0]/\partial{M_0} = \frac{1}{N_v}\left[0~0~1\right]^T$.

\bibliographystyle{IEEEtran}
\bibliography{bibliography}

\ifCLASSOPTIONcaptionsoff
  \newpage
\fi

\end{document}